\documentclass[prd,onecolumn]{revtex4}
\usepackage{dcolumn}
\usepackage{multirow}
\usepackage{graphicx}
\usepackage{amssymb}
\usepackage{bm}
\usepackage{color}
\usepackage{mathrsfs}
\usepackage{amsmath,amssymb,amsthm}

\begin{document}
\title{Observational constraints and differential diagnosis for cosmic evolutionary models}
\author{Deng Wang}
\email{Cstar@mail.nankai.edu.cn}
\affiliation{Theoretical Physics Division, Chern Institute of Mathematics, Nankai University,
Tianjin 300071, China}
\author{Xin-he Meng}
\email{xhm@nankai.edu.cn}
\affiliation{{Department of Physics, Nankai University, Tianjin 300071, P.R.China}\\
{State Key Lab of Theoretical Physics,
Institute of Theoretical Physics, CAS, Beijing 100080, P.R.China}}
\begin{abstract}
In this paper, we have proposed a plotting method based on the `` natural plotting rule '' (NPR) which can be used to distinguish different cosmological scenarios more efficiently and obtain more useful information. By using the NPR, we have avoided the blindness to use different diagnostics when discovering that some scenarios can be hardly differentiated from each other, and develop a logical line to adopt different diagnostics. As a concrete instance, we take this method based on the NPR to distinguish several Cardassian scenarios from the base cosmology scenario, and one from the other.
We place constraints on three Cardassian cosmological scenarios and their flat versions by utilizing the Type Ia supernovae (SNe Ia), baryonic acoustic oscillations (BAO), cosmic microwave background (CMB) radiation, observational Hubble parameter (OHD) data-sets as well as the single data point from the newest event GW150914, and discover that our results are more stringent than previous results for constraining the cosmological parameters of the Cardassian scenarios. We find that the flat original Cardassian (FOC) and original Cardassian (OC) scenarios can only be distinguished in the plane of $\{\Omega_m,S_3^{(1)}\}$ at the present epoch, however, if applying the NPR to plot hierarchically for these Cardassian scenarios in the plane of $\{S_3^{(1)},S_4^{(1)}\}$, we can obtain more detailed information and distinguish the two scenarios better than before. More importantly, from the planes of $\{S_4,S_4^{(2)}\}$, $\{S_5^{(1)},S_5^{(2)}\}$, $\{S_3^{(2)},S_4^{(2)}\}$, $\{\Omega_m,S_3^{(1)}\}$,$\{z,\omega\}$ $\{\epsilon(z),S_3^{(1)}\}$ and $\{z,Om\}$,
we dsicover that the flat modified polytropic Cardassian (FMPC) scenario can be directly removed from the possible candidates of dark energy phenomenon, since its evolutional behavior deviates from the base cosmology scenario too much.

\end{abstract}
\maketitle
\section{Introduction}
During the past few decades, a great deal of astronomical observations and gradually mounting evidence, including the cosmic microwave background (CMB) anisotropies, the observations of Type Ia supernovae (SNe Ia), the measurements of the baryonic acoustic oscillations (BAO) features, the abundance of galaxy clusters (AGC), the weak gravitational lensing (WGL) and so on, indicate that the universe is undergoing a phase of accelerated expansion \cite{1,2}. To explain the attractively and puzzlingly accelerated phenomenon, cosmologists proposed an additionally negative pressure fluid, dubbed dark energy. The most elegant candidate of dark energy is the so-called $\Lambda$CDM model, (which mainly contains two components, namely, cold dark matter and the cosmological constant $\Lambda$), since it is not only substantially concise but also well consistent with currently cosmological observations. However, this model faces two insurmountable problems, namely, the `` fine-tuning '' problem and `` coincidence '' problem \cite{3}. The former indicates that theoretical estimations for the vacuum energy density are many orders of magnitude larger than its observed value, i.e., the famous 120-orders-of-magnitude discrepancy that makes the vacuum explanation doubtful, while the latter implies why the amounts of the dark matter and dark energy are at the same order today since the scaling behavior of the energy densities are very different during the evolution of the universe by global fitting. Hence, the realistic nature of dark energy may not be the cosmological constant $\Lambda$ in the standard cosmological scenario. Based on this concern, theorists proposed a large number of alternatives in order to understand the dark energy phenomenon deeper, containing phantom \cite{4}, quintessence \cite{5,6,7,8,9,10,11,12}, bulk viscosity \cite{13,14,15,16,17,18,19}, Chaplygin gas \cite{20}, decaying vacuum \cite{21}, f(R) gravity \cite{21,22,23,24,25,26}, braneworld model \cite{27,28,29}, Einstein-Aether gravity \cite{31}, scalar-tensor theories of gravity \cite{32,33}, etc.

In this paper, we are aiming at investigating the so-called Cardassian universe which incarnates the hope that, it is a priori believable that Friedmann equation is modified in the very far infrared so that the universe begins to accelerate at late time. In Cardassian models \cite{34}, the universe is spatially flat and accelerating, and includes just two components, i.e., radiation and matter. Therefore, the Friedmann equation must be modified as follows:
\begin{equation}
H^2+\frac{k}{a^2}=\frac{g(\rho)}{3} \label{1.1},
\end{equation}
where $H$ denotes the Hubble parameter, $k$ the spatial curvature, $a$ the scalar factor of the universe, and $\rho$ the radiation and matter. Additionally, we have adopted the units of $8\pi G=c=1$ and the homogeneous and isotropic Friedmann-Robertson-Walker (FRW) metric
\begin{equation}
ds^2=-dt^2+a(t)^2[\frac{dr^2}{1-kr^2}+r^2(d\theta^2+sin^2\theta d\phi^2)] \label{1.2}.
\end{equation}

Generally speaking, to construct a reasonable Cardassian model, one must satisfy at least three basic requirements. First of all, the function $g(\rho)$ should reduce to the general form $\rho$ at early times so as to recover the thermal history of the base cosmology model and the scenario for the formation of the large scale structure (LSS). In the second place, $g(\rho)$ must adopt a different form at late epochs $z\sim\mathcal{O}(1)$ so as to drive an accelerated expansion. Finally, the sound speed $C^2_s$ of classical perturbations of the total cosmological fluid around the FRW solutions can not be negative, which reflects the classical solution of the accelerated expansion must be stable. During the past few years, several suitable Cardassian models satisfying the aforementioned three requirements have been proposed, containing the original Cardassian model (OC) \cite{34}, the modified polytropic Cardassian model (MPC) \cite{35}, the exponential Cardassian model (EC) \cite{36} as well as their flat version (FOC, FMPC, FEC), in which the spatial curvature is neglected.

Since so many Cardassian models have been proposed, it becomes substantially important and constructive to discriminate them from the $\Lambda$CDM model and one from the other in order to find better scenarios. So far several effective methods have appeared in the literature \cite{37,38}, i.e., the statefinder, the growth rate of matter perturbations, the $Om(z)$ diagnostic and the statefinder hierarchy. The statefinder which is only related to the third derivative of scale factor, is a robust and sensitive geometrical diagnostic of dark energy models. Thus, it can not distinguish some dark energy models well from each other. However, since the statefinder hierarchy proposed by Arabsalmani et al. \cite{39} is related to the higher order derivatives of the scale factor, it can break the degeneracy of different dark energy models better that the original statefinder. For a concrete instance, the original statefinder can not distinguish the new agegraphic dark energy model with different parameter values, but the statefinder hierarchy can discriminate them from each other well \cite{40}.

For this reason, in the following context, we would like to explore the evolutional behaviors and the relationships among different Cardassian models by using the statefinder hierarchy, the growth rate of matter perturbations and the $Om(z)$ diagnostic.

This paper is outlined in the following manner. In the next section, we make a brief review about the Cardassian models. In Sec. 3, we would like to constrain these models by adopting the SNe Ia, BAO, CMB, observational Hubble parameter data-sets (OHD) as well as the single data point from the newest event GW150914 \cite{41}. In Sec. 4, we briefly review the statefinder hierarchy, the growth rate of matter perturbations and the $Om(z)$ diagnostic. In Sec. 5, we discriminate the four models from each other and the base cosmology model by applying the mentioned-above three diagnostics. In the final section, we present the discussions and conclusions.

\section{The Cardassian models}
In paper \cite{34}, Freeze and Lewis proposed a new model as a possible explanation for the accelerated expansion of the universe, named Cardassian model (i.e., FOC), in which the Friedmann equation is modified without introducing the dark energy component. The Friedmann equation of this original flat Cardassian model can expressed as
\begin{equation}
H^2=\frac{\rho_m}{3}+B\rho_m^n \label{2.1},
\end{equation}
where $n$ is assumed to satisfy $n<2/3$ and $\rho_m$ just represents the matter component without considering the radiation contribution. The latter term is called Cardassian term which may exhibit that the universe as a (1+3)-dimensional brane is embedded in extra dimensions. The first term in Eq. \ref{2.1} will dominate at early epochs, so the equation becomes the same with the standard Friedmann equation. The second term comes to dominate recently, at the redshift $z\sim\mathcal{O}(1)$ indicated by the supernovae observations, so the universe tend to be accelerated at late times. In addition, it is worth noting that, if $B = 0$, it becomes the general Friedmann equation, but with only matter component. If $n=0$, it will become the same as the standard cosmological model. Subsequently, by adopting
\begin{equation}
\rho_m=\rho_{m0}(1+z)^3 \label{2.2},
\end{equation}
one can easily obtain
\begin{equation}
E^2(z)=\frac{H^2(z)}{H^2_0}=\Omega_{m0}(1+z)^3+(1-\Omega_{m0})(1+z)^{3n}  \label{2.3},
\end{equation}
where z denotes the redshift, $E(z)$ the dimensionless Hubble parameter, $\rho_{m0}$ the present-day value of the matter density, $H_0$ the present-day value of the Hubble parameter, and $\Omega_{m0}$ the present-day value of the matter density ratio parameter.

The flat modified polytropic Cardassian model (FMPC) \cite{35} is obtained by introducing an additional parameter $q$ into the original Cardassian model, which will reduce to the original model if $\beta=1$,
\begin{equation}
E^2(z)=\Omega_{m0}(1+z)^3+(1-\Omega_{m0}f_X(z)),  \label{2.3.1}
\end{equation}
where
\begin{equation}
f_X(z)=\frac{\Omega_{m0}}{1-\Omega_{m0}}(1+z)^3[(1+\frac{\Omega_{m0}^{-q}-1}{(1+z)^{3(1-n)q}})^{\frac{1}{q}}-1], \label{2.3.2}
\end{equation}
Note that, specially, if the parameters $\beta=1$ and $n=1$, $f_X(z)=1$ and this model will reduce to the base cosmology scenario.
Substitute Eq. \ref{2.3.2} into Eq. \ref{2.3.1}, one can conveniently get
\begin{equation}
E^2(z)=\Omega_{m0}(1+z)^3[1+(\Omega_{m0}^{-q}-1)(1+z)^{3q(n-1)}]^{\frac{1}{q}}  \label{2.3.3},
\end{equation}

Another interesting Cardassian model is the FEC model \cite{36}, and the corresponding Friedmann equation can be written as
\begin{equation}
H^2=\frac{g(\rho_m)}{3}=\frac{\rho_m}{3}\exp[(\frac{\rho_{card}}{\rho_m})^n]\label{2.4},
\end{equation}
where $\rho_{card}$ and $n$ are a characteristic constant energy density and a dimensionless constant, respectively. Note that we have neglected the contribution from the radiation. According to the third condition to construct a suitable Cardassian model, the accelerated expansion of the universe requires
\begin{equation}
3n(\frac{\rho_{card}}{\rho_{m0}})^n>1 \label{2.5}.
\end{equation}
Since the matter energy density during the evolutional process of the universe keeps conservative, one can obtain
\begin{equation}
\dot{\rho_m}+3H(\rho_m+p_m)=0 \label{2.6},
\end{equation}
where $p_m$ denotes the pressure of the matter component. Hence, the evolution of the matter also obeys the same equation, namely, Eq. \ref{2.2}. It is easy to be seen when $\rho$ is much larger than the characteristic energy density $\rho_{card}$, $g(\rho)\rightarrow\rho$. In terms of Eq. \ref{2.4} and \ref{2.6}, one can find that the effective pressure of the total fluid $p_t$ can be expressed in the following manner
\begin{equation}
p_t=\rho_m\frac{\partial{g(\rho_m)}}{\partial\rho_m}-g(\rho_m) \label{2.7}.
\end{equation}
Using the Eqs. \ref{2.4}, \ref{2.6} and \ref{2.7}, one can also derive the background evolution equation of the EC model as follows:
\begin{equation}
E^2(z)=\Omega_{m0}(1+z)^3\exp[-(1+z)^{-3n}\ln\Omega_{m0}] \label{2.8}.
\end{equation}
Subsequently, we would not like to depict the left Cardassian models one by one which will be studied in the present work, and we summarize all the models in Table. \ref{tab1}
\begin{table}[h!]
\centering
\begin{tabular}{|ll|l|l|l}
\hline
&$g(\rho_m)$           & Model       &$E^2(z)$    \\
\hline
&$\rho_m[1+(\frac{\rho_m}{\rho_{card}})^{n-1}]$      &FOC    &$\Omega_{m0}(1+z)^3+(1-\Omega_{m0})(1+z)^{3n}$        \\
&                                                     &OC    &$\Omega_{m0}(1+z)^3+(1-\Omega_{m0}-\Omega_{k0})(1+z)^{3n}+\Omega_{k0}(1+z)^2$                                                       \\
& $\rho_m[1+(\frac{\rho_m}{\rho_{card}})^{q(n-1)}]^{\frac{1}{q}}$       &FMPC      &$\Omega_{m0}(1+z)^3[1+(\Omega_{m0}^{-q}-1)(1+z)^{3q(n-1)}]^{\frac{1}{q}}$   \\
&        &MPC      &$\Omega_{m0}(1+z)^3[1+((1-\Omega_{k0})^q\Omega_{m0}^{-q}-1)(1+z)^{3q(n-1)}]^{\frac{1}{q}}+\Omega_{k0}(1+z)^2$     \\
&$\rho_m\exp[(\frac{\rho_m}{\rho_{card}})^{-n}]$                &FEC    &$\Omega_{m0}(1+z)^3\exp[-(1+z)^{-3n}\ln\Omega_{m0}]$     \\
&                                                               &EC   &$\Omega_{m0}(1+z)^3\exp[-(1+z)^{-3n}(\ln\Omega_{m0}-\ln(1-\Omega_{k0}))]+\Omega_{k0}(1+z)^2$      \\

\hline
\end{tabular}
\caption{Six Cardassian models with different functions $g(\rho_m)$ at the late epoch of the universe.}
\label{tab1}
\end{table}

\section{The statistical analysis}
\subsection{Type Ia supernovae}
As is well known, in modern astronomy, theoretically, SNe Ia can be used as the standard candles to probe the expansion history of our universe, since all the SNes Ia almost explode at the same mass ($M\approx-19.3\pm0.3$). In the present situation, we take the Union 2.1 data-sets without systematic errors for fitting, consisting of 580 data points. To perform the so-called $\chi^2$ statistics, the theoretical distance modulus for a supernovae at redshift z, given a set of model parameters $\theta$, is defined as
\begin{equation}
\mu_{t}(z_i)=m-M=5\log_{10}D_L(z_i)+25\label{3.1},
\end{equation}
where $m$ is the apparent magnitude, $M$ the absolute magnitude and $D_L(z_i)$ the luminosity distance at a given redshift $z_i$ in units of megaparsecs,
\begin{equation}
D_L=(1+z)\int^z_0\frac{dz'}{E(z';\theta)}\label{3.2},
\end{equation}
where $E(z';\theta)$ denotes the dimensionless Hubble parameter. Subsequently, the corresponding $\chi_S^2$ for the SNe Ia observations can be expressed as
\begin{equation}
\chi^2_S=\sum^{580}_{i=1}[\frac{\mu_{o}(z_i)-\mu_{t}(z_i;\theta)}{\sigma_i}]^2 \label{3.3},
\end{equation}
where $\mu_{o}$ and $\sigma_i$ denote the observed value and the corresponding 1¦Ò error of the distance modulus, respectively, at a given redshift $z_i$.
\subsection{Baryonic acoustic oscillations}
The location of the BAO peak provides a general ruler, with a constant comoving scale at different redshifts during the whole evolution history of the universe, since the baryons on the scales about 150 Mpc are non-relativistic after short recombination, the location of the peak in the comoving coordinate would not change. Subsequently, we would like to take the BAO data-sets appearing in \cite{42}, and use the dsitance parameter $\mathcal{A}$ to measure the BAO peak in the distribution of the Sloan Digital Sky Survey (SDSS) luminous red galaxies, and the distance parameter $\mathcal{A}$ can be defined as
\begin{equation}
\mathcal{A}=\sqrt{\Omega_{m0}}E(z_i)^{-\frac{1}{3}}[\frac{1}{z_i}\int^{z_i}_0\frac{dz'}{E(z')}]^{\frac{2}{3}} \label{3.4}.
\end{equation}
The corresponding $\chi^2$ for the BAO data-sets is
\begin{equation}
\chi^2_{B}=\sum^6_{i=1}[\frac{\mathcal{A}_{o}(z_i)-\mathcal{A}_{t}(z_i;\theta)}{\sigma_{\mathcal{A}}}]^2 \label{3.5},
\end{equation}
where $\mathcal{A}_{o}$ denotes the observed value of the distance parameter $\mathcal{A}$, $\mathcal{A}_{t}$ the theoretical value of the distance parameter $\mathcal{A}$ and $\sigma_{\mathcal{A}}$ the corresponding 1¦Ò error of the distance parameter $\mathcal{A}$, at a given redshift $z_i$.
\subsection{Cosmic microwave background}
As an important and effective supplement, we would like to adopt the CMB shift parameter to calculate the combined analysis in order to make the constraints more strictly. It is noteworthy that the CMB shift parameter might be the least independent model parameter that could be extracted from the CMB data-sets and it can be defined as follows \cite{43}
\begin{equation}
\mathcal{R}=\sqrt{\Omega_{m0}}\int^{z_C}_0\frac{dz'}{E(z')} \label{3.6},
\end{equation}
where $z_C$ is the redshift of recombination. The seven-year Wilkinson Microwave Anisotropy Probe (WMAP) results \cite{44} have shown the value of $z_C$ as $z_C=1091.3$ independent of the dark energy model and the shift parameter $\mathcal{R}=1.725\pm0.018$. The corresponding $\chi^2$ for the CMB observations can be expressed as follows
\begin{equation}
\chi^2_{C}(\theta)=[\frac{\mathcal{R}(\theta)-1.725}{0.018}]^2 \label{3.7}.
\end{equation}
\subsection{Observational Hubble parameter}
The OHD can be used to constrain the cosmological model parameters since they are obtained from model-independent direct observations and there is no need to integrate over the redshift $z$ so as to drop some useful information. Up to now, there are two main methods to measure the OHD in the literature, i.e., galaxy differential age and radial BAO size methods \cite{45}. We refer the readers to the paper \cite{46}, in order to obtain more useful information. Usually, the corresponding $\chi^2$ for the OHD can be defined as
\begin{equation}
\chi^2_{H}=\sum^{29}_{i=1}[\frac{H_0E(z_i)-H_{o}(z_i)}{\sigma_i}]^2 \label{3.8},
\end{equation}
where $H_{o}(z_i)$ denotes the observed value of the OHD.
\subsection{Gravitational wave data}
The detection by Abbott et al. \cite{41} of gravitational radiation from a pair of merging massive black holes, $36^{+5}_{-4}$ $M_{\odot}$ and $29^{+4}_{-4}$ $M_{\odot}$, has ushered in a new era of the multi-messenger astronomy. This gravitational source lies at the luminosity distance of $410^{+160}_{-180}$ Mpc corresponding to the redshift $z=0.09^{+0.03}_{-0.04}$. In the present situation, we would like to transform the single data point to the SNe Ia data point in order to constrain the Cardassian models as well, calculating out the corresponding distance modulus $38.0639^{+0.7155}_{-1.2553}$. which never appears in the previous literature. Although the quality of the data is not very good, we believe strongly that the future gravitational-wave data-sets will open a new and powerful window for new physics. For simplicity, in the following context, we would like to denote the statistical contribution from the single gravitational wave data point as $\chi^2_G$.
\begin{figure}
\centering
\includegraphics[scale=0.5]{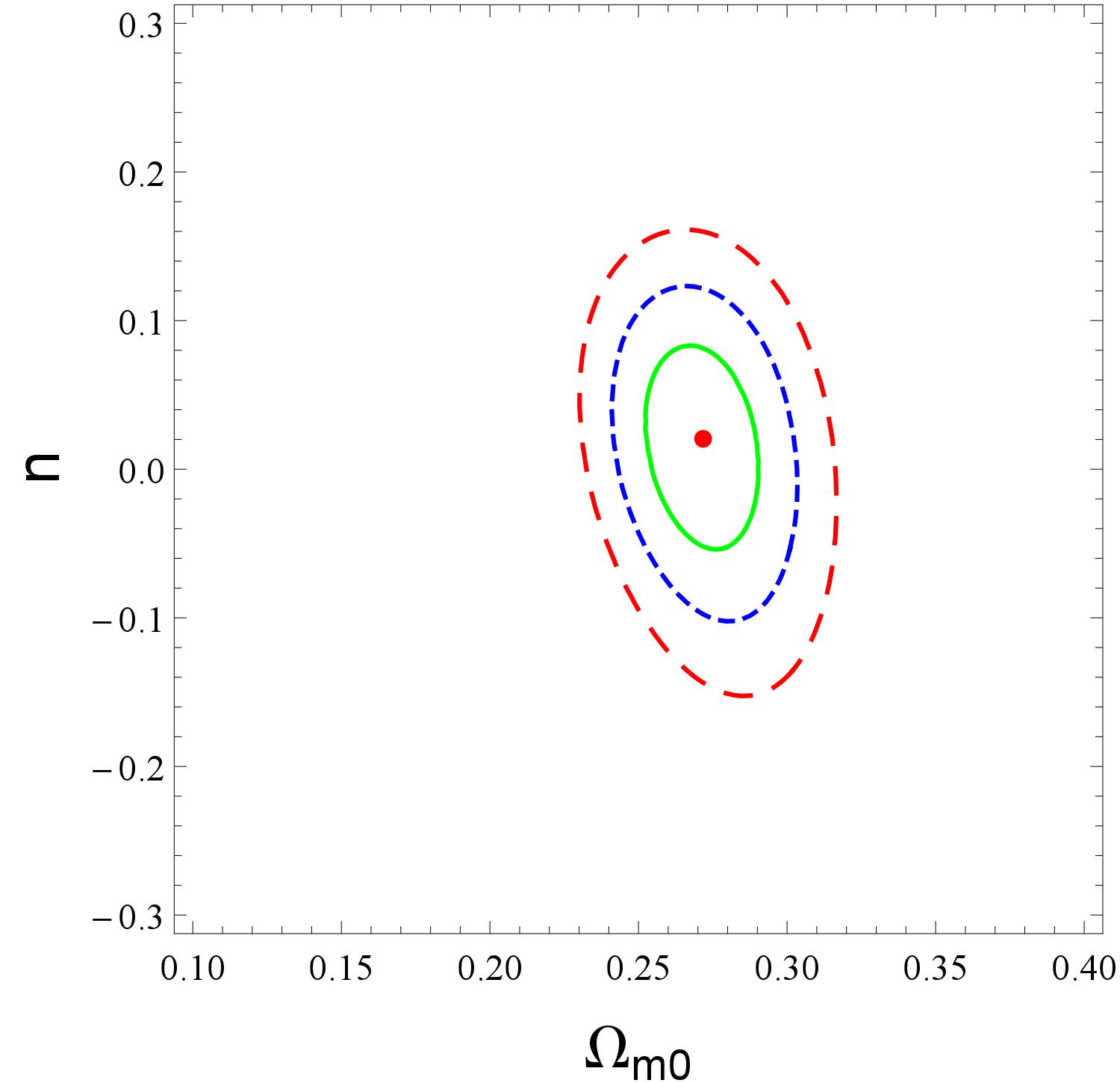}
\caption{1$\sigma$, 2$\sigma$ and 3$\sigma$ confidence levels for parameter pair ($\omega_0$, $n$) of the FOC model, constrained by SNe Ia, BAO, CMB, OHD and the gravitational wave data-sets. The best fitting value is shown as a dot which corresponds to $\Omega_{m0}=0.270862$, $n=0.0169296$.}
\label{f1}
\end{figure}
\begin{figure}
\centering
\includegraphics[scale=0.5]{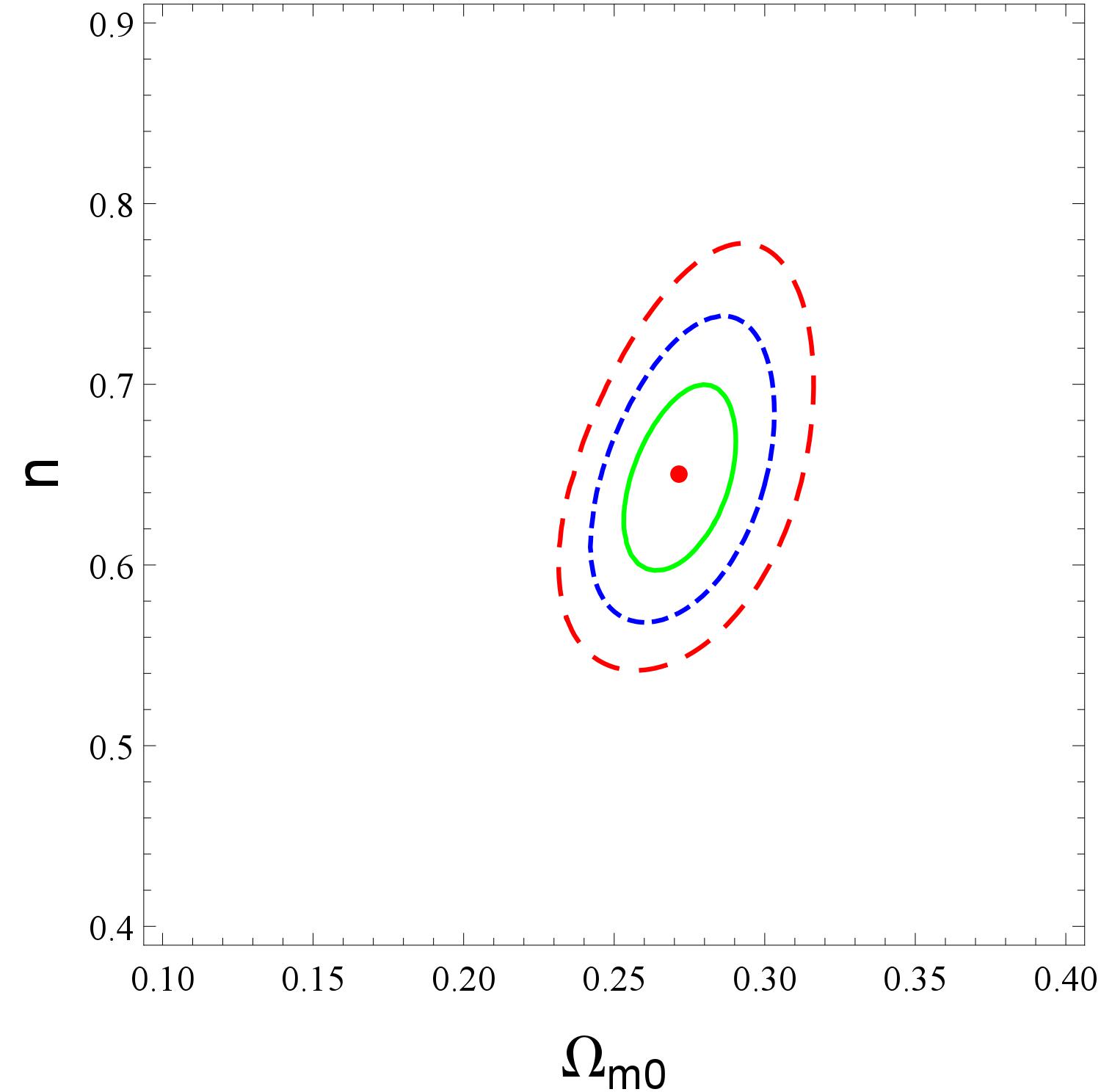}
\caption{1$\sigma$, 2$\sigma$ and 3$\sigma$ confidence levels for parameter pair ($\omega_0$, $n$) of the FEC model, constrained by SNe Ia, BAO, CMB, OHD and the gravitational wave data-sets. The best fitting value is shown as a dot which corresponds to $\Omega_{m0}=0.271232$, $n=0.645559$.}
\label{f2}
\end{figure}

The next step is to compute the joint constraints from SNe Ia, BAO, CMB, OHD and the gravitational wave data-sets, and the corresponding $\tilde{\chi}^2$ can be defined as follows
\begin{equation}
\tilde{\chi}^2={\chi}^2_{S}+\chi^2_{B}+\chi^2_{C}+{\chi}^2_{H}+\chi^2_G \label{3.9}.
\end{equation}
The minimum values $\chi^2_{min}$ of the derived $\tilde{\chi}^2$ and the best fitting values of the model parameters constrained by the SNe Ia, BAO, CMB, OHD and the gravitational wave data-sets, are listed in Table. \ref{tab2}. At the same time, we just perform the likelihood distributions of the model parameters ($\Omega_{m0}$, $n$) of the joint constraints $\tilde{\chi}^2$ from the aforementioned data-sets for the models FOC and FEC.
In addition, we find that our results are more stringent than previous results for constraining the cosmological parameters of the Cardassian cosmological models \cite{47}. In the following context, we would like to use the the best fitting values of the parameters to distinguish the Cardassian models from the base cosmology model and one from the other.
\begin{table}[h!]
\centering
\begin{tabular}{|ll|l|l|l}
\hline
          &Model       &$\chi^2_{min}$    &Best fitting values   \\
\hline
          &FOC         &578.437           &$\Omega_{m0}=0.270862$, $n=0.0169296$                               \\
          &OC          &578.334           &$\Omega_{m0}=0.271578$, $n=-0.0178564$, $\Omega_{k0}=0.0213974$      \\
          &FMPC        &578.479           &$\Omega_{m0}=0.270968$, $n=0.0413404$, $q=1.04192$                    \\
          &MPC         &578.416           &$\Omega_{m0}=0.27346$, $n=0.138889$, $q=0.0531985$, $\Omega_{k0}=1.48286$ \\
          &FEC         &580.558           &$\Omega_{m0}=0.271232$, $n=0.645559$                                       \\
          &EC          &578.559           &$\Omega_{m0}=0.267105$, $n=0.446996$, $\Omega_{k0}=-0.236004$               \\
\hline
\end{tabular}
\caption{The best fitting values of the model parameters for the different Cardassian cosmological models by adopting the joint constraints from the SNe Ia, BAO, CMB, OHD and the gravitational wave data-sets.}
\label{tab2}
\end{table}
\section{The geometrical diagnostics}
In the previous literature, there are three main diagnostics to discriminate various kinds of cosmological models from each other, namely, the statefinder hierarchy, the growth rate of matter perturbations and the $Om(z)$ diagnostic. Subsequently, we will make a brief review about the three null diagnostics one by one.
\subsection{The statefinder hierarchy}
As well known, the primary aim of the statefinder hierarchy is to distinguish the dynamical dark energy models from the $\Lambda$CDM model. Afterwards, since it not only contains higher derivatives of the scale factor $d^na/dt^n$ ($n\geqslant2$), but also has a favorable property that all the members of the statefinder hierarchy can be expressed in terms of the deceleration parameter $q$ (consequently the matter density parameter $\Omega_{m}$), so that it may discriminate the dark energy models better. Then, we Taylor-expand the scale factor $a(t)$ around the present epoch $t_0$ in the following manner:
\begin{equation}
\frac{a(t)}{a_0}=1+\sum\limits_{n=1}^\infty\frac{A_n(t_0)}{n!}[H_0(t-t_0)]^n \label{4.1},
\end{equation}
where
\begin{equation}
A_n=\frac{a^{(n)}}{aH^n} \label{4.2},
\end{equation}
where $a^{(n)}=d^na/dt^n$ and $n\in N$. Note that a great deal of letters of the alphabet have been taken to represent different derivatives of the scale factor $a(t)$. More specifically, $q=-A_2$ denotes the familiar deceleration parameter, $A_3$ the original statefinder `` $r$ '', $A_4$ the snap `` $s$ '' and $A_5$ the lerk `` $l$ ''. For the base cosmology model, one can easily obtain
\begin{eqnarray}
  A_2&=&1-\frac{3}{2}\Omega_m \label{4.3}, \\
  A_3&=&1 \label{4.4},  \\
  A_4&=&1-\frac{3^2}{2}\Omega_m \label{4.5},  \\
  A_5&=&1+3\Omega_m+\frac{3^3}{2}\Omega^2_m  \label{4.6},\qquad ...,
\end{eqnarray}
where $\Omega_m=2(1+q)/3$ and $\Omega_m=\Omega_{m0}(1+z)^3/E^2(z)$ for the base cosmology model. Subsequently, the zeroth-order statefinder hierarchy $S_n$ can be defined as follows
\begin{eqnarray}
  S_2&=&A_2+\frac{3}{2}\Omega_m \label{4.7}, \\
  S_3&=&A_3 \label{4.8},   \\
  S_4&=&A_4+\frac{3^2}{2}\Omega_m \label{4.9},   \\
  S_5&=&A_5-3\Omega_m-\frac{3^3}{2}\Omega^2_m  \label{4.10},\qquad ....
\end{eqnarray}
It is easy to be proved that, for the base cosmology model, the zeroth-order statefinder hierarchy $S_n$ can be expressed as
\begin{equation}
S_n\mid_{\Lambda CDM}=1 \label{4.11}.
\end{equation}
It is worth noting that the above equations just give out an excellent mull diagnostic for the base cosmology model, since the equalities will be violated by other dark energy models. Furthermore, when $n\geqslant3$, one can define a series of statefinders in the following manner:
\begin{eqnarray}
  S^{(1)}_3&=&S_3 \label{4.12}, \\
  S^{(1)}_4&=&A_4+3(1+q) \label{4.13},   \\
  S^{(1)}_5&=&A_5-2(4+3q)(1+q)  \label{4.14},\qquad....
\end{eqnarray}
These statefinders have the completely same property with $S_n$, i.e., remaining pegged at unity in the process of the evolution of the universe for the base cosmology model:
\begin{equation}
S^{(1)}_n\mid_{\Lambda CDM}=1 \label{4.15}.
\end{equation}
Hence, one could get an appealing and important property, i.e., $\{S_n,S_n^{(1)}\}\mid_{\Lambda CDM}=1$, which also can be acted as an effective tool to discriminate other dark energy models form the base cosmology model. In addition, the second-order statefinder hierarchy can be constructed from $S_n^{(1)}$
as follows:
\begin{equation}
S^{(2)}_n=\frac{S^{(1)}_n-1}{3(q-\frac{1}{2})} \label{4.16}.
\end{equation}
For the the base cosmology scenario, it is obvious that $\{S_n,S^{(2)}_n \}=\{1,0 \}$, $\{S^{(1)}_n,S^{(2)}_n \}=\{1,0 \}$. As for the evolving dark energy models, one will obtain different results in order to be distinguished from the base cosmology scenario better. According to the original paper \cite{39}, $\omega$CDM, Chaplygin gas (CG), and DGP model have been discriminated from each other and the base cosmology scenario. In our previous work \cite{48}, four time-dependent dark energy models also can be distinguished well from the base cosmology scenario and one from the other in terms of $\{S_3^{(1)},S_4^{(1)}\}$, $\{S_3^{(2)},S_4^{(2)}\}$, etc.
\subsection{The growth rate of perturbations}
The growth rate of perturbations can be regarded as an important supplement for the statefinder hierarchy, and the fractional growth parameter $\epsilon(z)$ can be defined as follows \cite{49,50}
\begin{equation}
\epsilon(z)=\frac{f(z)}{f_{\Lambda CDM}(z)} \label{4.17},
\end{equation}
where
\begin{equation}
f(z)=\Omega_m(z)^{\gamma(z)} \label{4.18},
\end{equation}
representing the growth rate of the linearized density perturbations, and
\begin{equation}
\gamma(z)=\frac{3}{5-\frac{\omega}{1-\omega}}+\frac{3}{125}\frac{(1-\omega)(1-1.5\omega)}{(1-1.2\omega)^3}[1-\Omega_m(z)]+\mathcal{O}[(1-\Omega_m(z))]^2 \label{4.19}.
\end{equation}
The equation of state parameter $\omega$ is either a constant, or varies slowly with time. Nonetheless, it is not the case in extended theories of gravities (ETGs) where the perturbation growth contains information which is just complementary to that contained in the expansion history. For the base cosmology scenario, it is not difficult to discover that
\begin{equation}
\epsilon(z)\mid_{\Lambda CDM}=1 \label{4.20}.
\end{equation}
Subsequently, we would like to combine the statefinder hierarchy with the fractional growth parameter $\epsilon(z)$ to define a composite null diagnostic (CND): $\{\epsilon(z),S_n\}$, $\{\epsilon(z),S_n^{(1)}\}$ or $\{\epsilon(z),S_n^{(2)}\}$. For the base cosmology scenario, $\gamma\simeq0.55$ and $\epsilon=1$ \cite{51,52}, thus, $\{\epsilon(z),S_n\}=\{1,1\}$.
\subsection{The $Om(z)$ diagnostic}
The $Om(z)$ diagnostic is another substantially useful method to distinguish various kinds of dark energy models from the $\Lambda$CDM model and one from the other, which can be defined as follows
\begin{equation}
Om(x)=\frac{E^2(x)-1}{x^3-1} \label{4.21},
\end{equation}
where $E(x)=H(x)/H_0$ and $x=1/a=1+z$. Similarly, neglecting the radiation contribution at low redshifts, for the $\Lambda$CDM model, one can easily get
\begin{equation}
E^2(x)=\Omega_{m0}x^3+(1-\Omega_{m0})  \label{4.22}.
\end{equation}
Substituting Eq. \ref{4.21} into Eq. \ref{4.22}, one can obtain
\begin{equation}
Om(x)\mid_{\Lambda CDM}=\Omega_{m0} \label{4.23}.
\end{equation}
One can easily find that the $Om(z)$ diagnostic also gives out an effective null test for the base cosmology scenario, and for other dynamical dark energy models, the $Om(z)$ diagnostics are expected to give out distinctive results.
\section{The differential diagnosis by using the statefinder hierarchy, the composite null diagnostic and the $Om(z)$ diagnostic}
In this section, we will adopt the statefinder hierarchy, the composite null diagnostic and the $Om(z)$ diagnostic to distinguish the above-mentioned Cardassian models from each other and from the base cosmology scenario. According to the original paper \cite{39}, the parameters $q$, $A_3$, $A_4$ and $A_5$ can be expressed as follows
\begin{eqnarray}
  q&=&(1+z)\frac{1}{E}\frac{dE}{dz}-1 \label{5.1}, \\
  A_3&=&(1+z)\frac{1}{E^2}\frac{d[E^2(1+q)]}{dz}-3q-2 \label{5.2},   \\
  A_4&=&-(1+z)\frac{1}{E^3}\frac{d[E^3(2+3q+A_3)]}{dz}+4A_3+3q(q+4)+6 \label{5.3},   \\
  A_5&=&-(1+z)\frac{1}{E^4}\frac{d[E^4(A_4-4A_3-3q(q+4)-6)]}{dz}+5A_4-10A_3(q+2)-30q(q+2)-24. \nonumber\\
                                                                                             \label{5.4}
\end{eqnarray}

\begin{figure}
\centering
\includegraphics[scale=0.5]{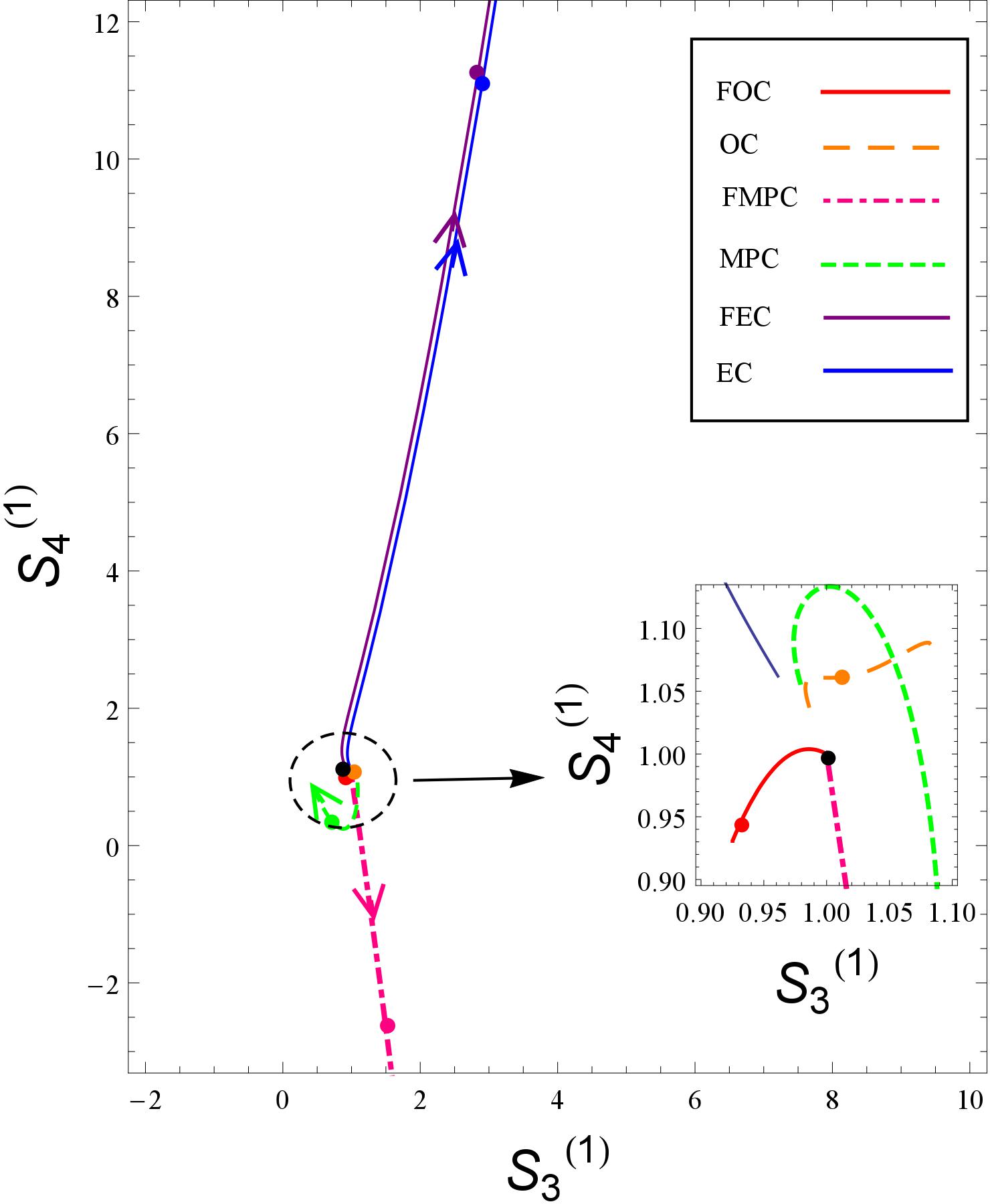}
\caption{The statefinder $\{S_3^{(1)},S_4^{(1)}\}$ plane.  The red (solid) line, the orange (long-dashed) line, the pink (dash-dotted) line, the green (dashed) line, the purple (solid) line and the blue (solid) line correspond to model FOC, model OC, model FMPC, model MPC, model FEC and model EC, respectively. The arrows imply the evolutional direction with respect to the cosmic time and the present epoch in different models is shown as a dot. The black dot corresponding to the fixed point $\{1,1\}$ represents the base cosmology scenario.}
\label{1}
\end{figure}

In the present situation, first of all, we adopt the first-order statefinder hierarchy $\{S_3^{(1)},S_4^{(1)}\}$ to distinguish the six Cardassian cosmological models from the base cosmology scenario and one from the other. In Figure. \ref{1}, it is easy to be seen that, if we adopt the relatively small plotting scale, the models FOC and OC can not be distinguished from each other and the base cosmology model at the present epoch, the models FEC and EC share the same evolution trend in the far future and they also can not be distinguished from each other at the present epoch, but they can be discriminated from the base cosmology model and other Cardassian models at the present stage. In addition, the models FMPC and MPC can be distinguished well from each other and other Cardassian models as well as the base cosmology scenario at the present epoch. It is noteworthy that if we plot in relatively large scale, we find that the models FOC and OC can be well distinguished from each other and the base cosmology scenario. Thus, when using the geometrical diagnostics to discriminate different dark energy models, how to make a appropriate choice for the plotting scale is a substantially key point. From our point of view, the figure in which one happens to put the present epochs of all the dark energy models is a good and natural one in practice. Consequently, in the following context, we would like to obey the `` natural plotting rule '' (NPR).
\begin{figure}
\centering
\includegraphics[scale=0.5]{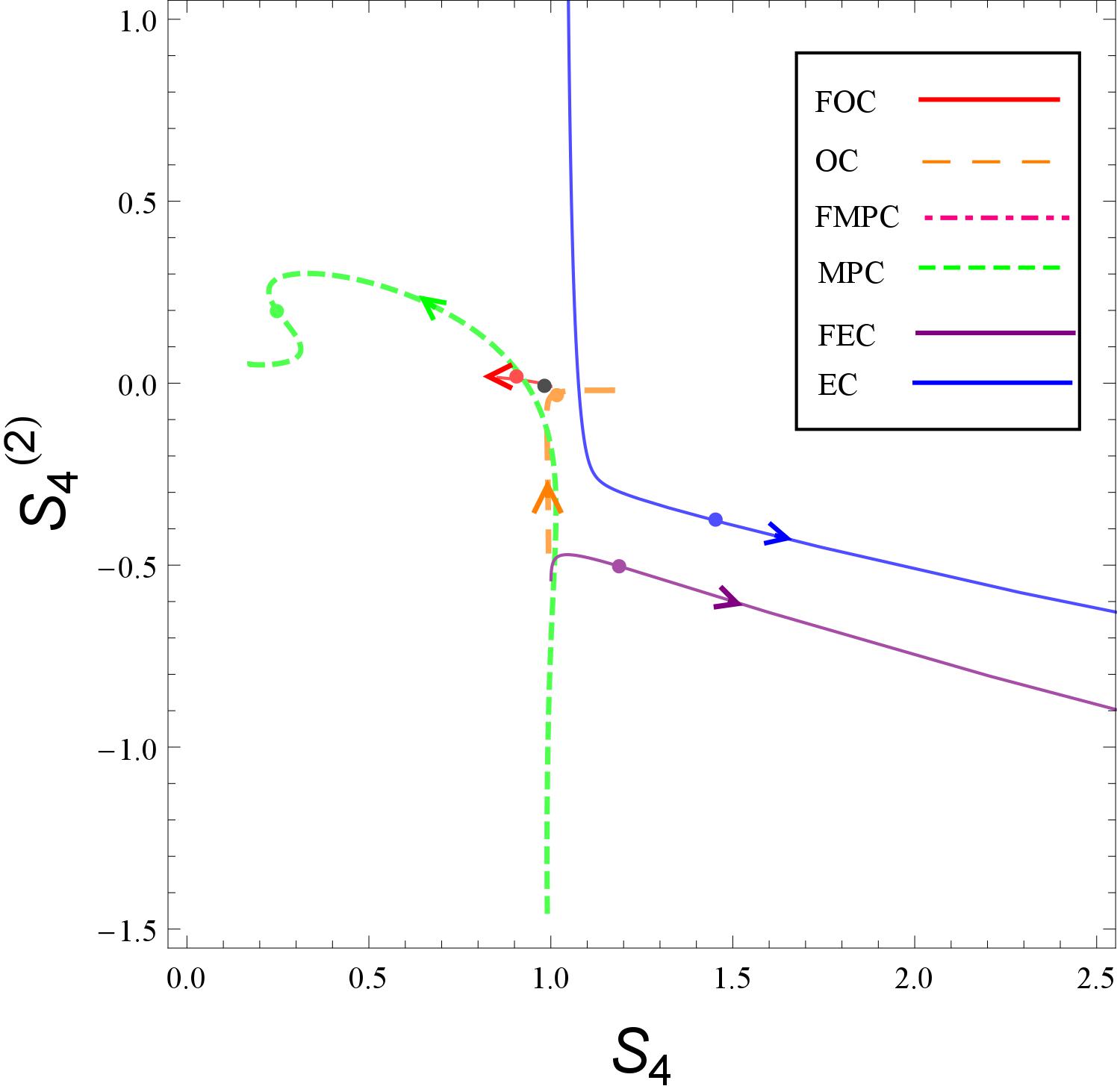}
\caption{The statefinder $\{S_3^{(1)},S_4^{(1)}\}$ plane. Note that we have used the NPR to plot for the left five Cardassian scenarios apart from the scenario FMPC here. The arrows imply the evolutional direction with respect to the cosmic time and the present epoch in different models is shown as a dot. The black dot corresponding to the fixed point $\{1,1\}$ represents the base cosmology scenario.}
\label{2}
\end{figure}
\begin{figure}
\centering
\includegraphics[scale=0.5]{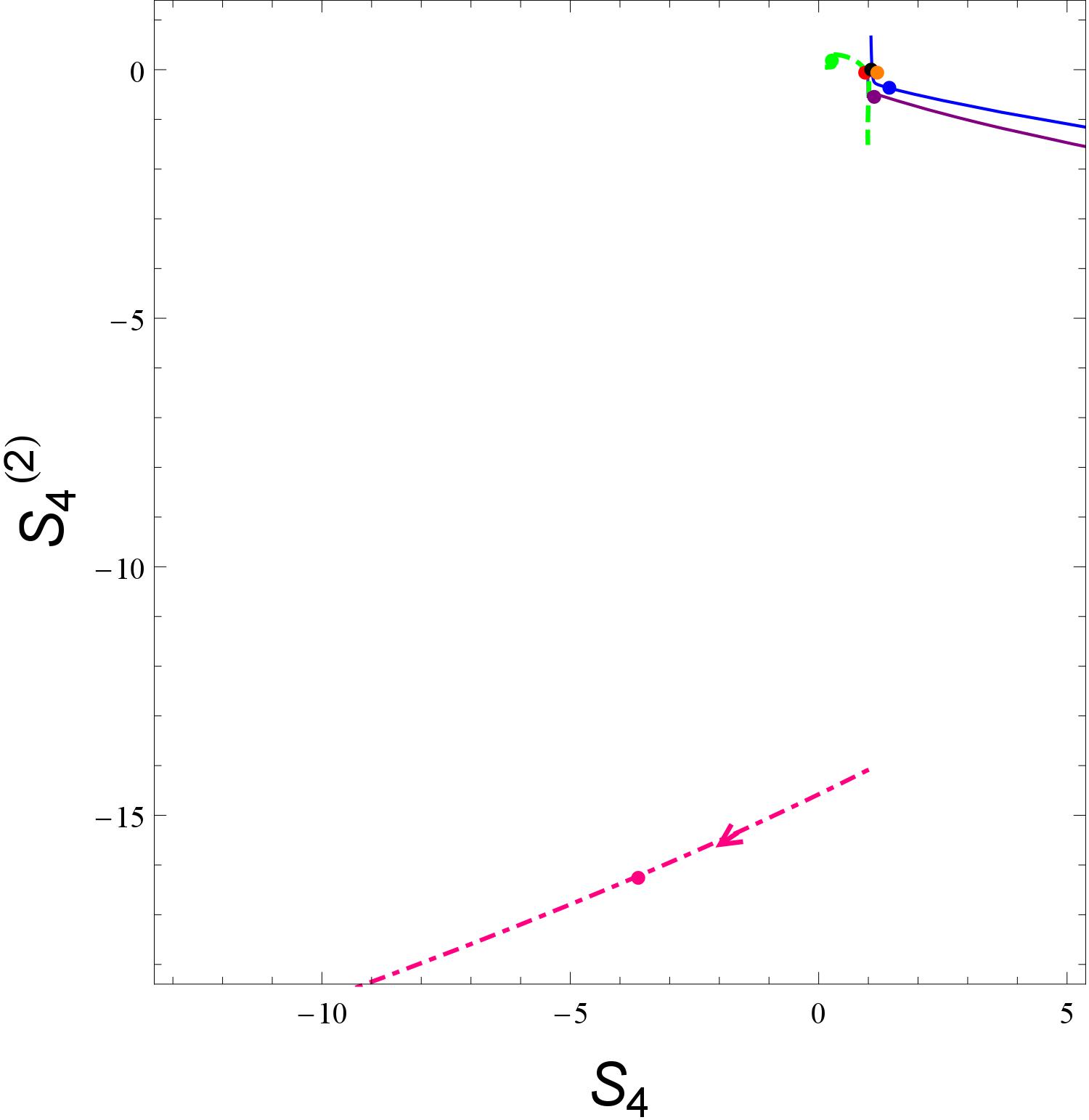}
\caption{The statefinder $\{S_4,S_4^{(2)}\}$ plane. The red (solid) line, the orange (long-dashed) line, the pink (dash-dotted) line, the green (dashed) line, the purple (solid) line and the blue (solid) line correspond to model FOC, model OC, model FMPC, model MPC, model FEC and model EC, respectively. The arrows imply the evolutional direction with respect to the cosmic time and the present epoch in different models is shown as a dot. The black dot corresponding to the fixed point $\{1,1\}$ represents the base cosmology scenario.}
\label{3}
\end{figure}
\begin{figure}
\centering
\includegraphics[scale=0.5]{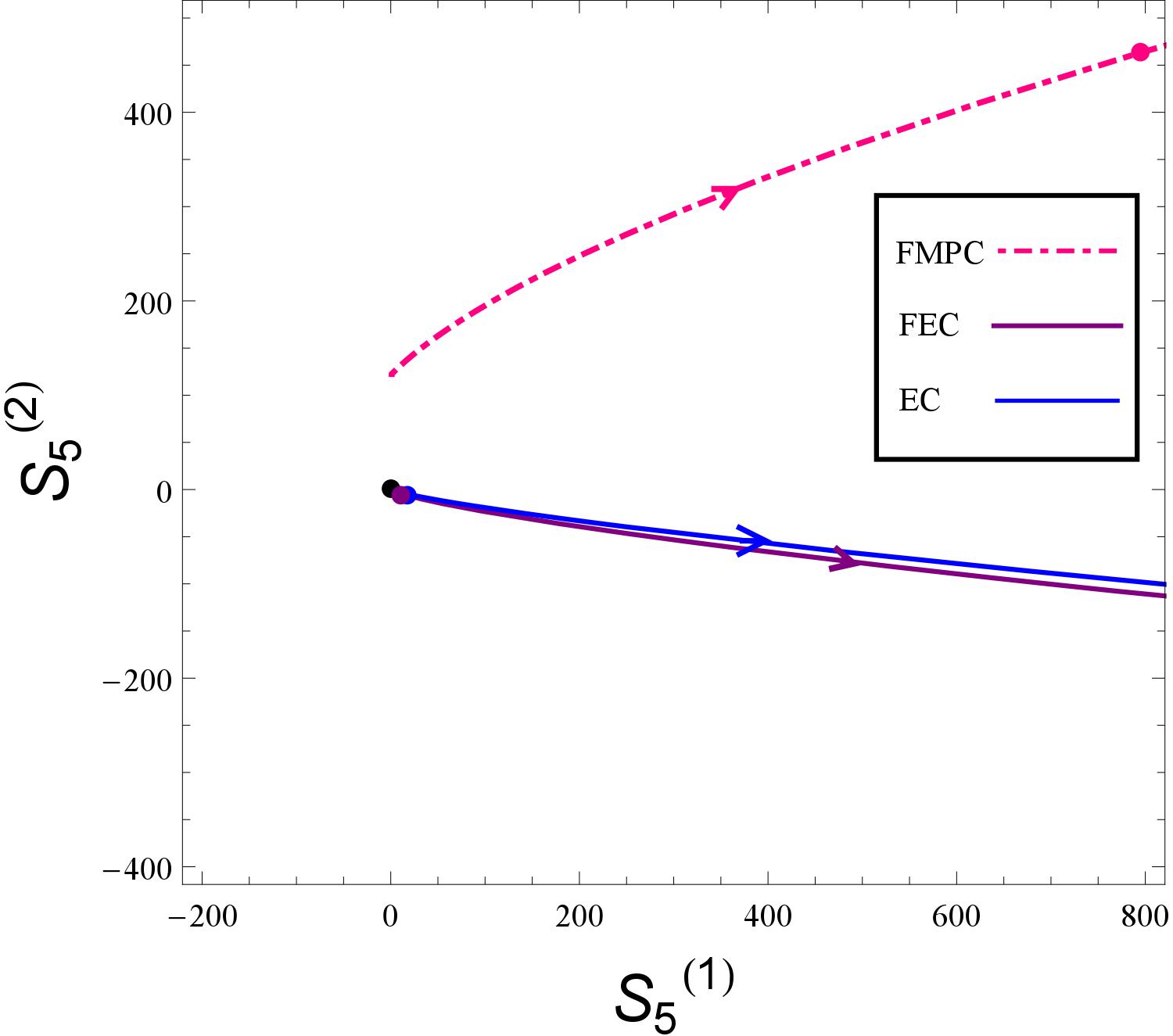}
\caption{The statefinder $\{S_5^{(1)},S_5^{(2)}\}$ plane. The pink (dash-dotted) line, the purple (solid) line and the blue (solid) line correspond to model FMPC, model FEC and model EC, respectively. The arrows imply the evolutional direction with respect to the cosmic time and the present epoch in different models is shown as a dot. The black dot corresponding to the fixed point $\{1,1\}$ represents the base cosmology scenario.}
\label{4}
\end{figure}
\begin{figure}
\centering
\includegraphics[scale=0.5]{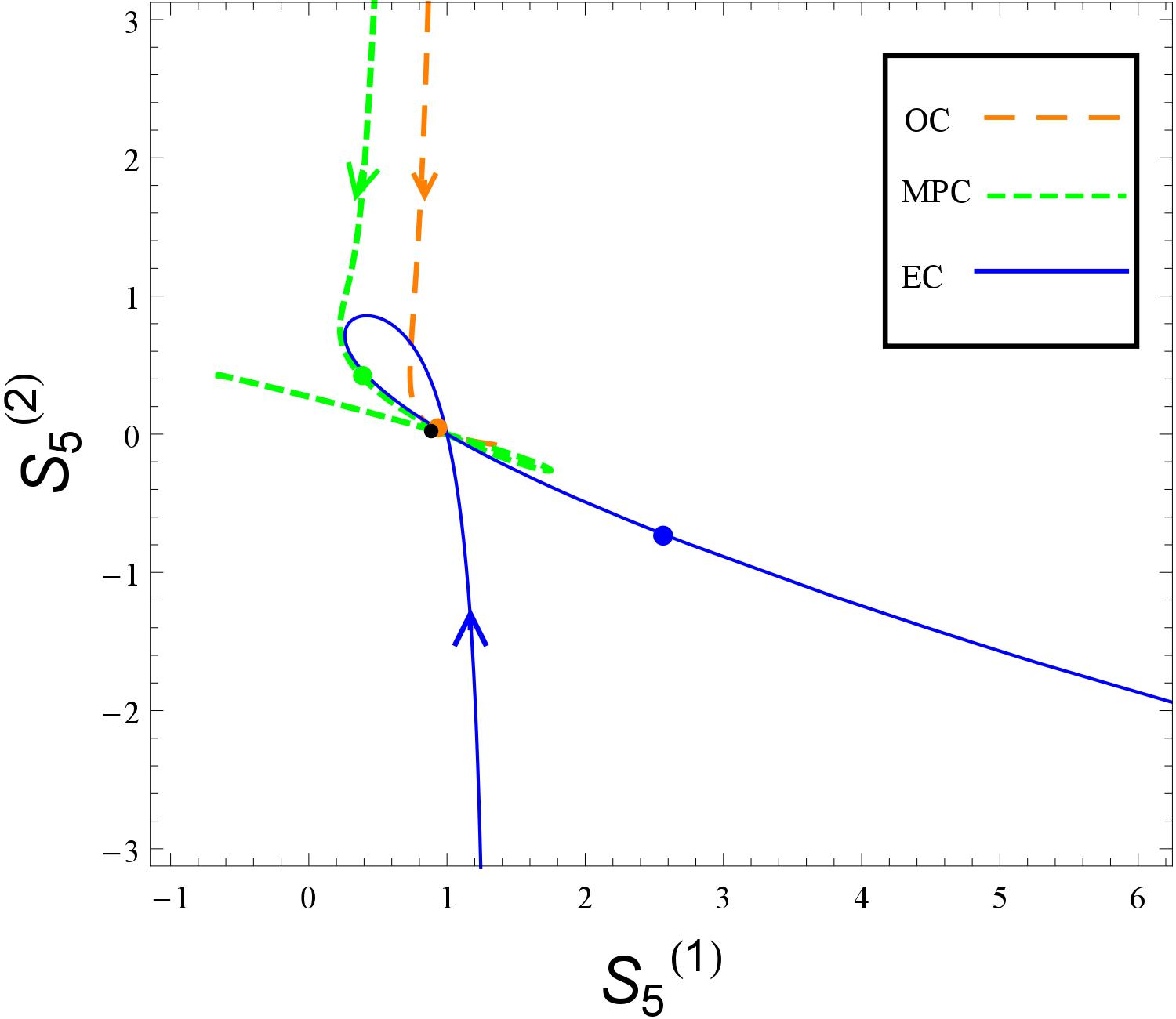}
\caption{The statefinder $\{S_5^{(1)},S_5^{(2)}\}$ plane. Note that we have used the NPR to plot for models OC, MPC and EC. The arrows imply the evolutional direction with respect to the cosmic time and the present epoch in different models is shown as a dot. The black dot corresponding to the fixed point $\{1,1\}$ represents the base cosmology scenario.}
\label{5}
\end{figure}
\begin{figure}
\centering
\includegraphics[scale=0.5]{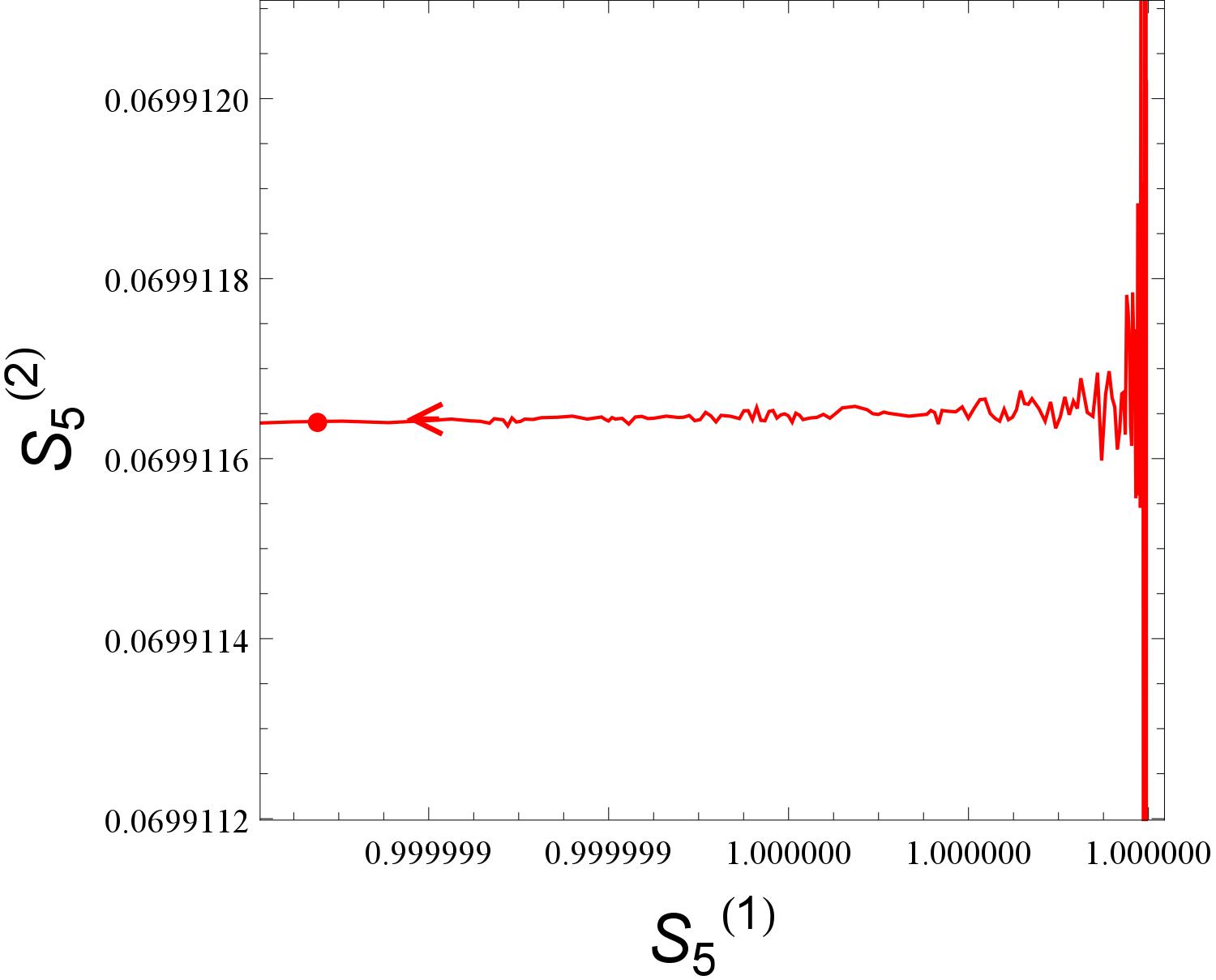}
\caption{The statefinder $\{S_5^{(1)},S_5^{(2)}\}$ plane. Note that we have used the NPR to plot for model FOC alone. The arrows imply the evolutional direction with respect to the cosmic time and the present epoch in different models is shown as a dot.}
\label{6}
\end{figure}
\begin{figure}
\centering
\includegraphics[scale=0.5]{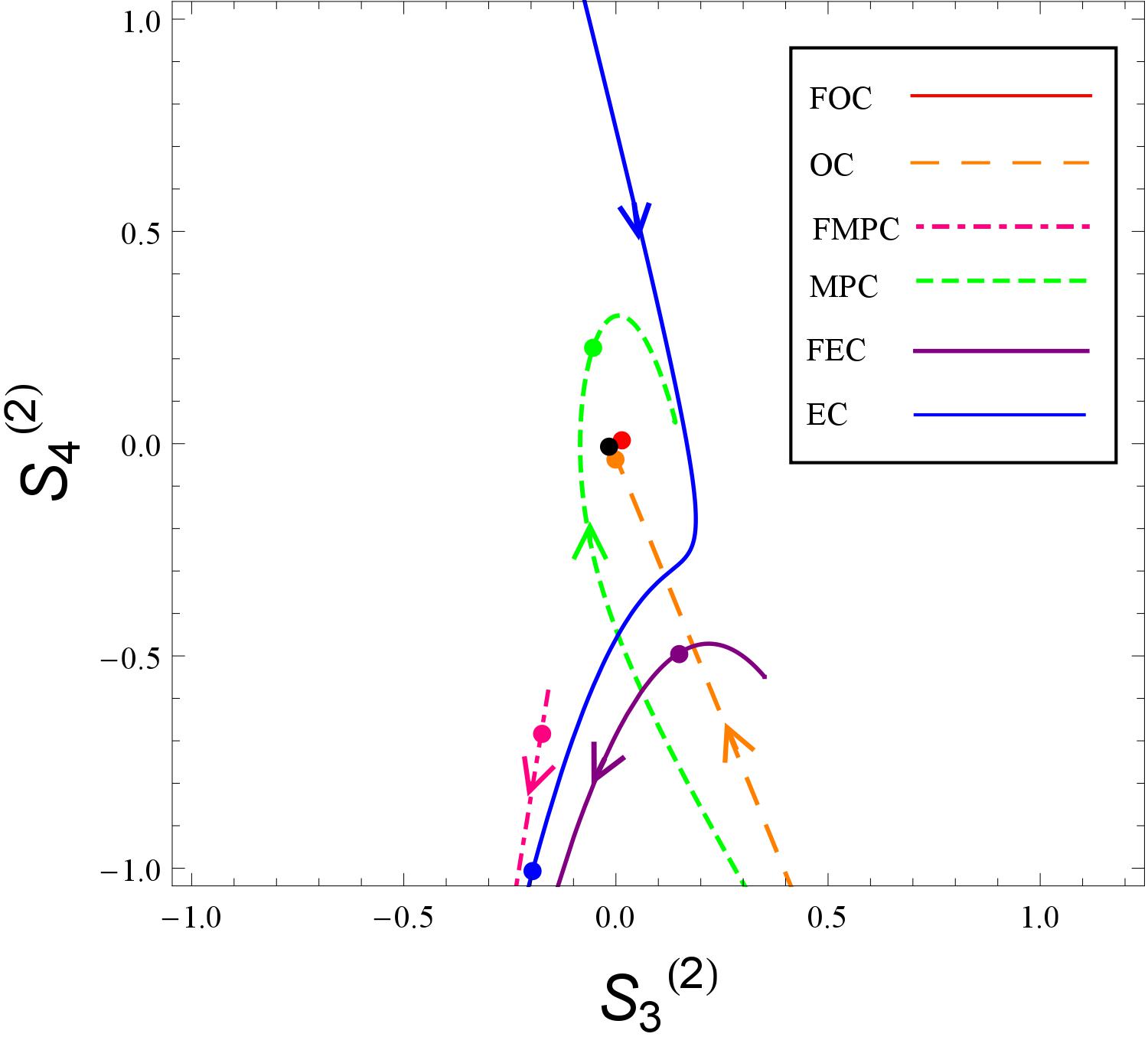}
\caption{The statefinder $\{S_3^{(2)},S_4^{(2)}\}$ plane. The red (solid) line, the orange (long-dashed) line, the pink (dash-dotted) line, the green (dashed) line, the purple (solid) line and the blue (solid) line correspond to model FOC, model OC, model FMPC, model MPC, model FEC and model EC, respectively. The arrows imply the evolutional direction with respect to the cosmic time and the present epoch in different models is shown as a dot. The black dot corresponding to the fixed point $\{1,1\}$ represents the base cosmology scenario.}
\label{7}
\end{figure}
\begin{figure}
\centering
\includegraphics[scale=0.5]{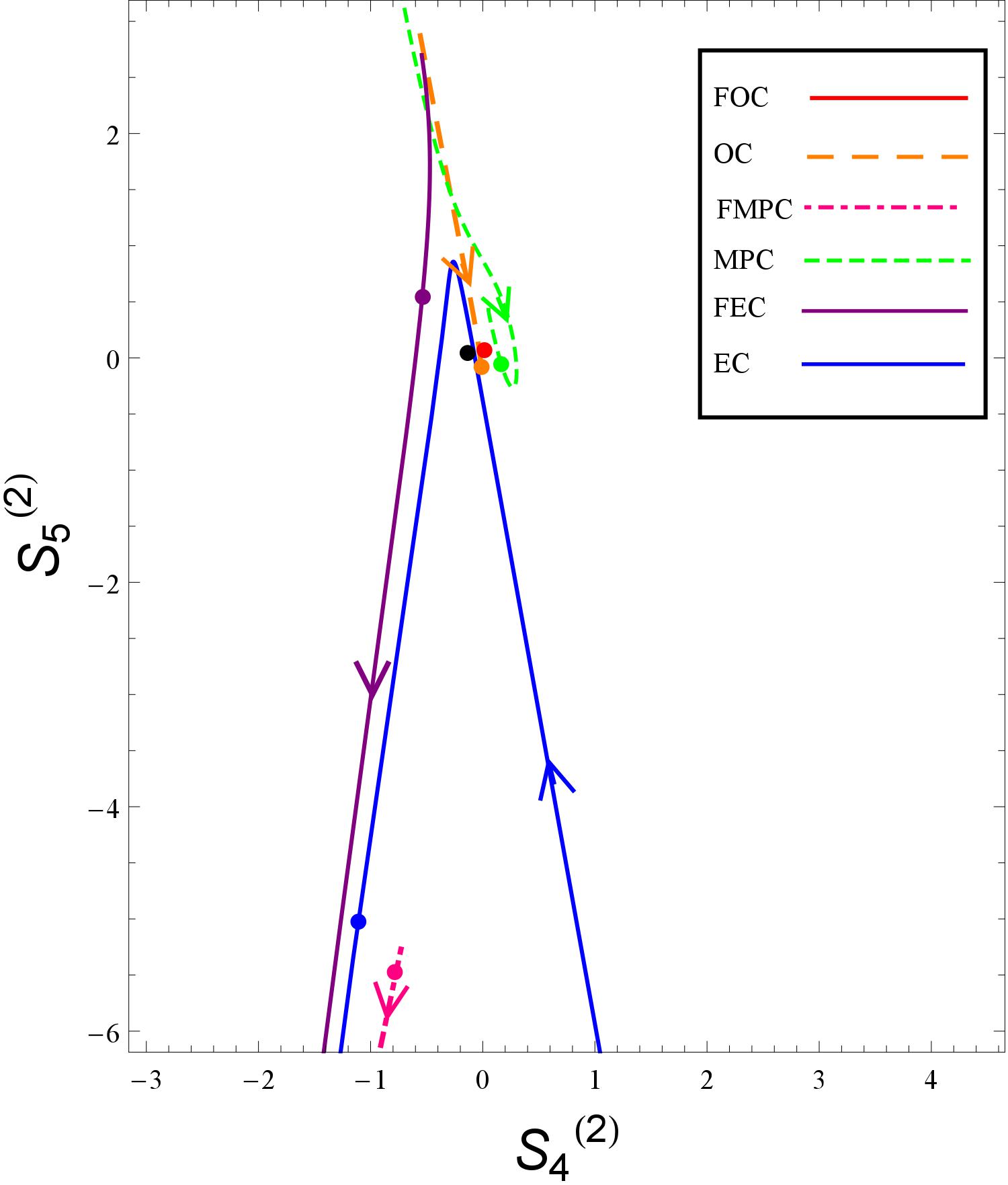}
\caption{The statefinder $\{S_4^{(2)},S_5^{(2)}\}$ plane. The red (solid) line, the orange (long-dashed) line, the pink (dash-dotted) line, the green (dashed) line, the purple (solid) line and the blue (solid) line correspond to model FOC, model OC, model FMPC, model MPC, model FEC and model EC, respectively. The arrows imply the evolutional direction with respect to the cosmic time and the present epoch in different models is shown as a dot. The black dot corresponding to the fixed point $\{1,1\}$ represents the base cosmology scenario.}
\label{8}
\end{figure}
\begin{figure}
\centering
\includegraphics[scale=0.5]{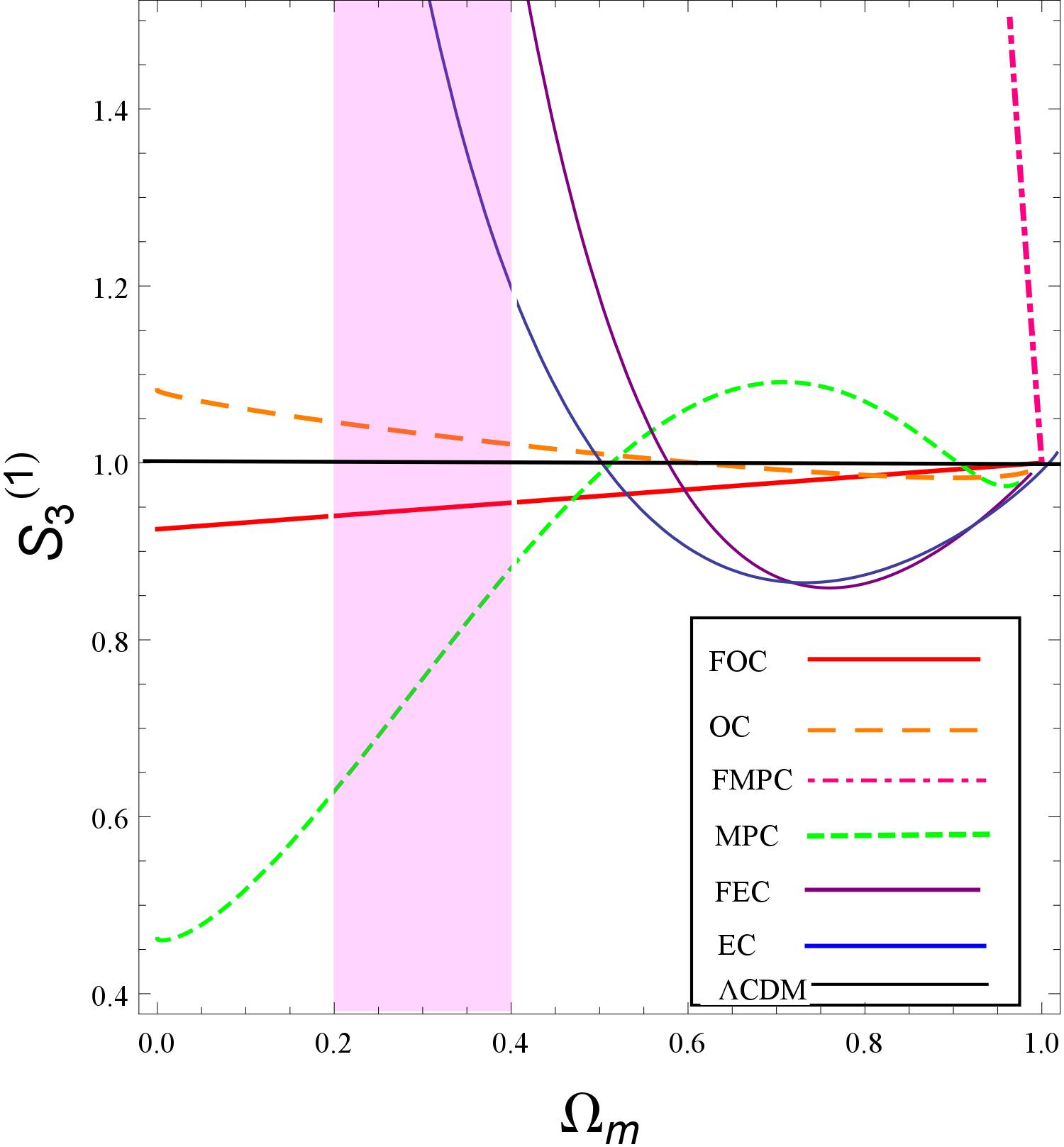}
\caption{The relation between the statefinder $S_3^{(1)}$ and the matter density parameter $\Omega_m$. The black (horizontal) line, the red (solid) line, the orange (long-dashed) line, the pink (dash-dotted) line, the green (dashed) line, the purple (solid) line and the blue (solid) line correspond to the base cosmology model, model FOC, model OC, model FMPC, model MPC, model FEC and model EC, respectively. The vertical band centered at $\Omega_{m0}=0.3$ roughly represents the present epoch.}
\label{9}
\end{figure}
\begin{figure}
\centering
\includegraphics[scale=0.5]{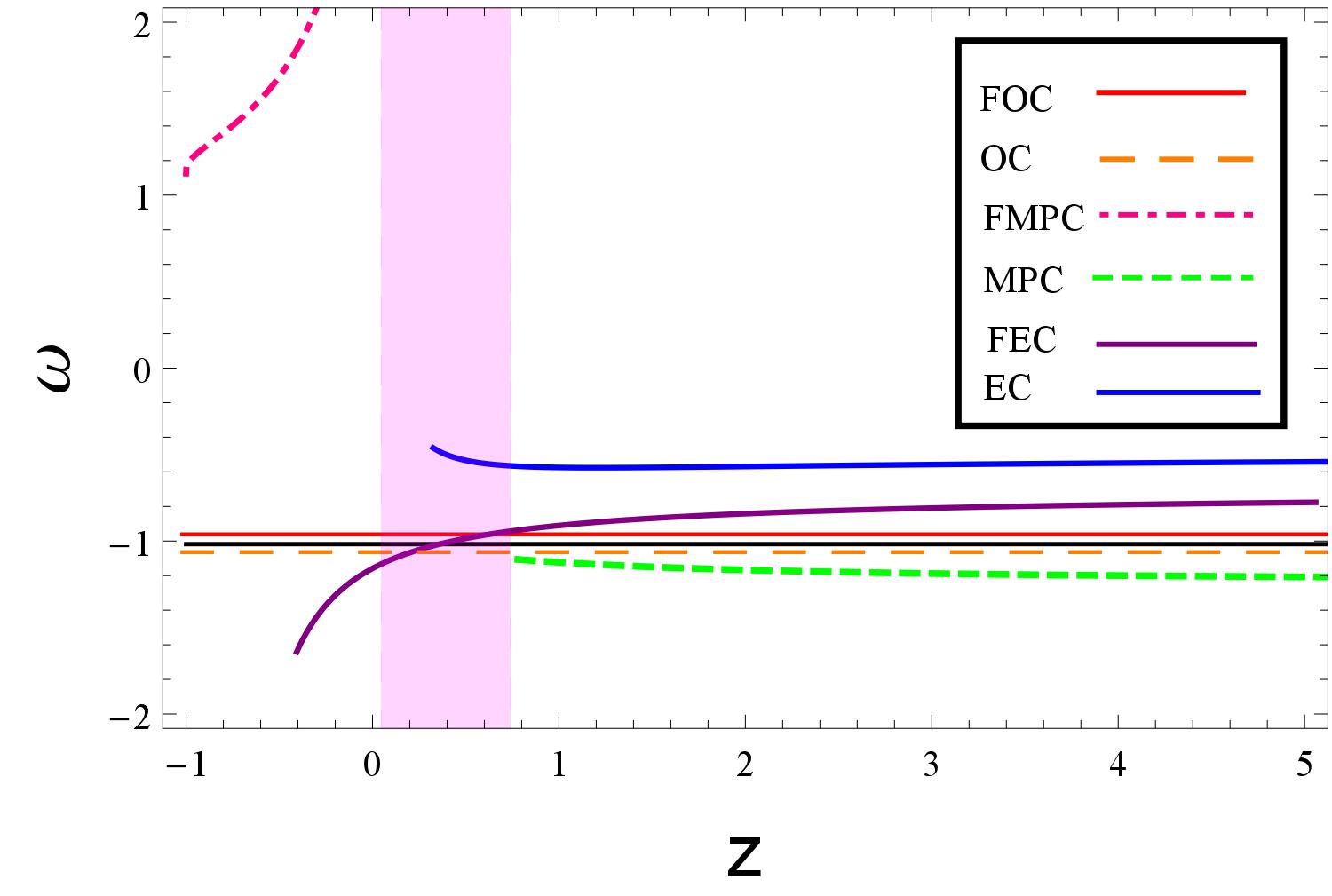}
\caption{The relation between the equation of state parameter $\omega$ and the redshift $z$. The black (horizontal) line, the red (solid) line, the orange (long-dashed) line, the pink (dash-dotted) line, the green (dashed) line, the purple (solid) line and the blue (solid) line correspond to the base cosmology model, model FOC, model OC, model FMPC, model MPC, model FEC and model EC, respectively. The vertical band roughly represents the present epoch.}
\label{10}
\end{figure}
\begin{figure}
\centering
\includegraphics[scale=0.5]{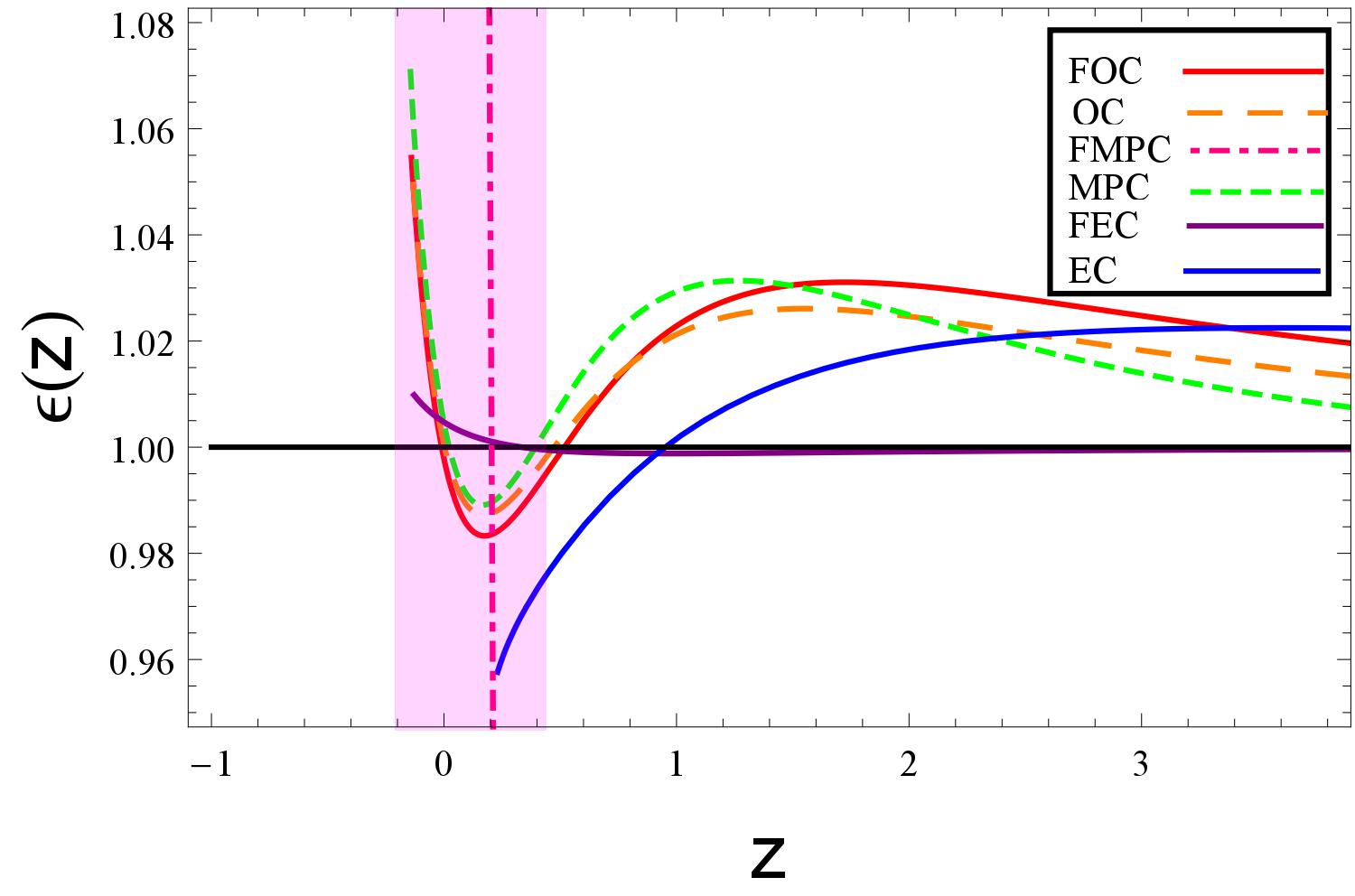}
\caption{The relation between the fraction growth parameter $\epsilon(z)$ and the redshift $z$. The black (horizontal) line, the red (solid) line, the orange (long-dashed) line, the pink (dash-dotted) line, the green (dashed) line, the purple (solid) line and the blue (solid) line correspond to the base cosmology model, model FOC, model OC, model FMPC, model MPC, model FEC and model EC, respectively. The vertical band roughly represents the present epoch.}
\label{11}
\end{figure}
\begin{figure}
\centering
\includegraphics[scale=0.5]{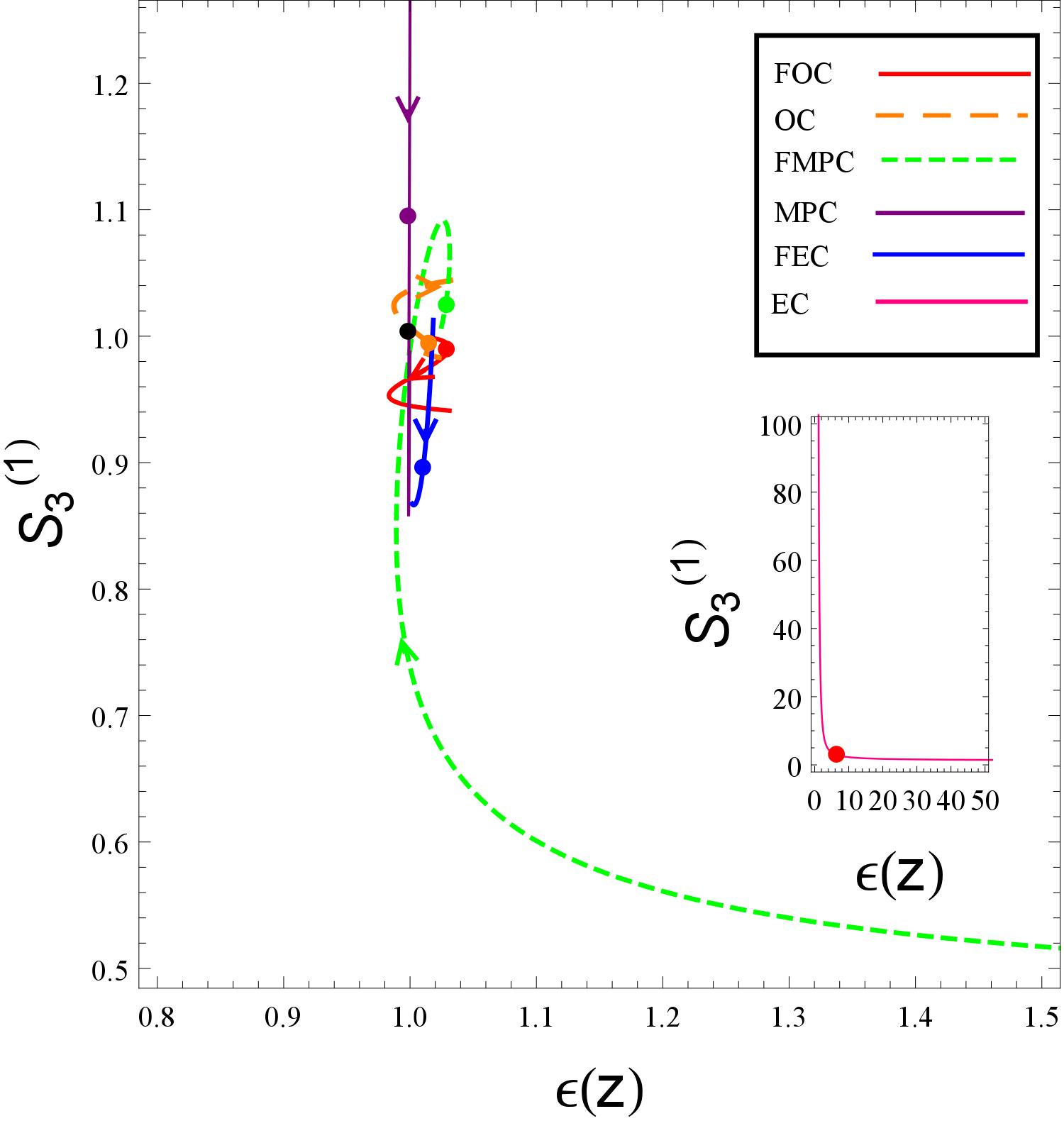}
\caption{The CND $\{\epsilon(z),S_3^{(1)}\}$ plane. The black (horizontal) line, the red (solid) line, the orange (long-dashed) line, the pink (dash-dotted) line, the green (dashed) line, the purple (solid) line and the blue (solid) line correspond to the base cosmology model, model FOC, model OC, model FMPC, model MPC, model FEC and model EC, respectively. The black dot corresponding to the fixed point $\{1,1\}$ represents the base cosmology scenario.}
\label{12}
\end{figure}
\begin{figure}
\centering
\includegraphics[scale=0.5]{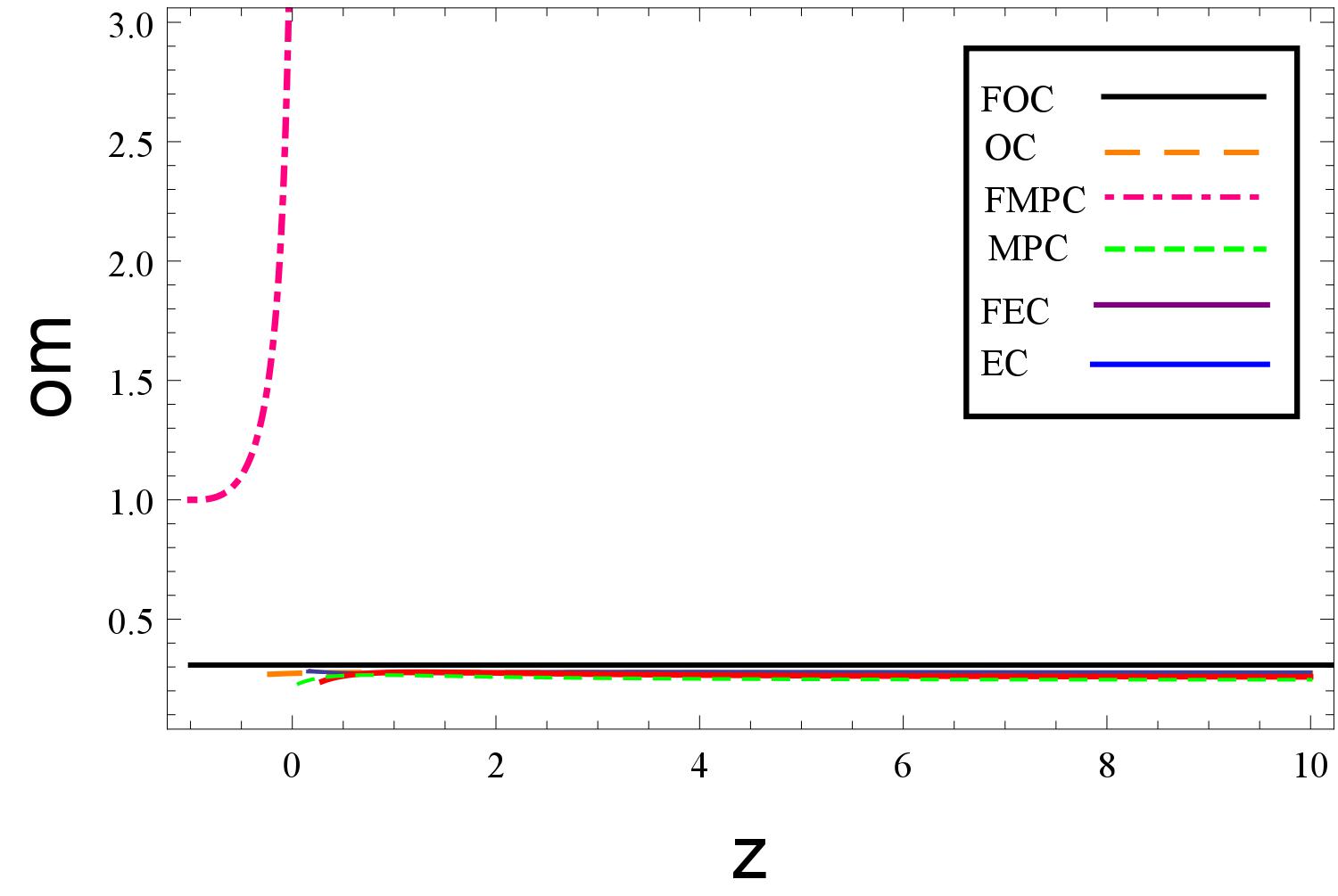}
\caption{The relation between $Om(z)$ and the redshift $z$. The black (horizontal) line, the red (solid) line, the orange (long-dashed) line, the pink (dash-dotted) line, the green (dashed) line, the purple (solid) line and the blue (solid) line correspond to the base cosmology model, model FOC, model OC, model FMPC, model MPC, model FEC and model EC, respectively.}
\label{13}
\end{figure}
\begin{figure}
\centering
\includegraphics[scale=0.5]{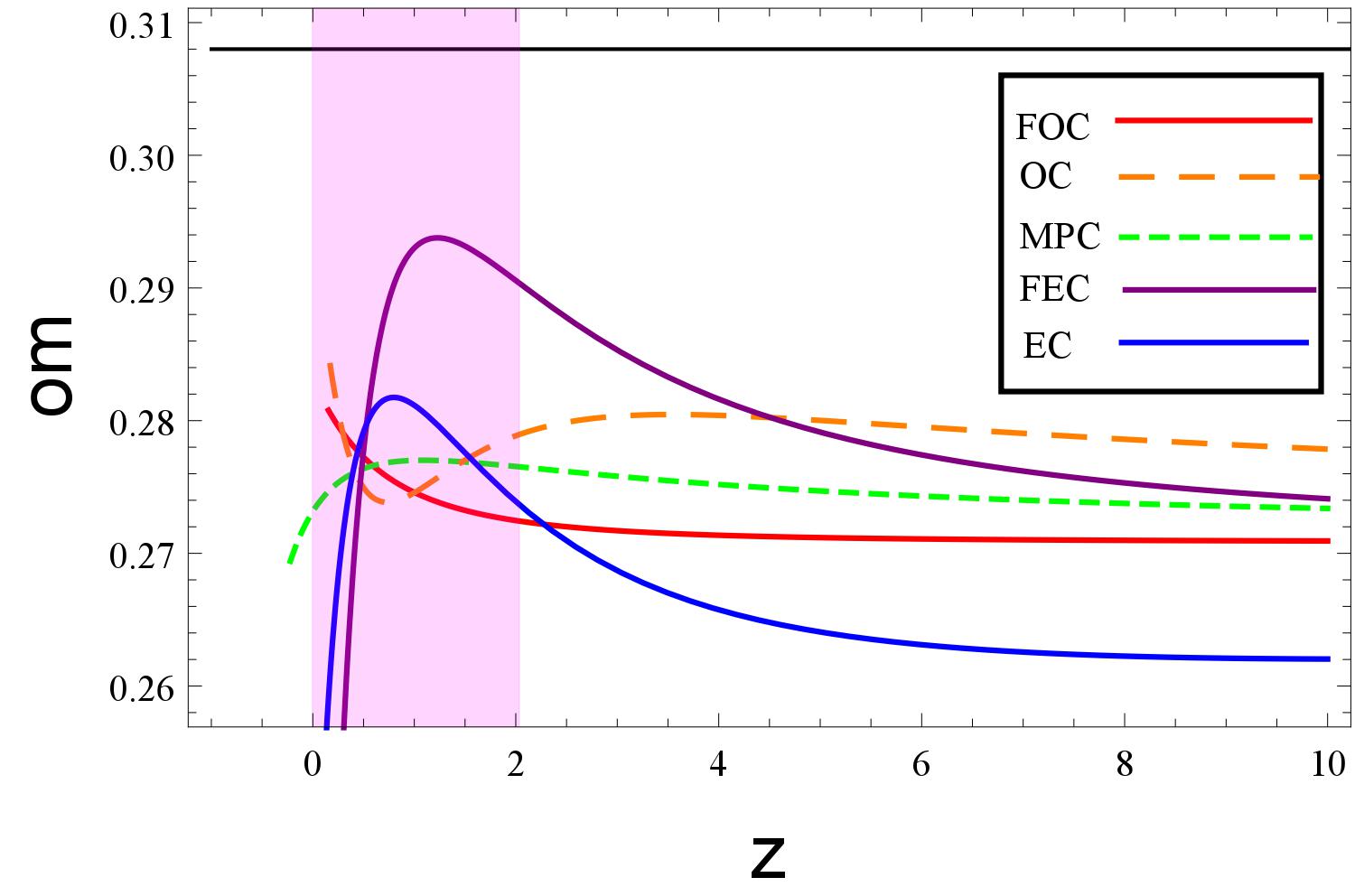}
\caption{The relation between $Om(z)$ and the redshift $z$. Note that we have used the NPR to plot for the left five Cardassian scenarios apart from the scenario FMPC here. The black (horizontal) line, the red (solid) line, the orange (long-dashed) line, the green (dashed) line, the purple (solid) line and the blue (solid) line correspond to the base cosmology model, model FOC, model OC, model MPC, model FEC and model EC, respectively. The vertical band roughly represents the present epoch.}
\label{14}
\end{figure}

In Figure. \ref{2}, we find that the models MPC, FEC and EC are distinguished obviously from each other and the base cosmology scenario at the present stage, but the models FOC and OC can be hardly discriminated from the base cosmology scenario at the present epoch. Note that the model FMPC does not appear in Figure. \ref{2}, since we have chosen the relatively large plotting scale. Therefore, if we choose a relatively small plotting scale, the model FMPC can be given out
clearly. In Figure. \ref{3}, it is obvious that the model FMPC deviates the base cosmology scenario and other Cardassian models so much that this model can be directly removed from the possible candidates of the dark energy phenomenon. Apart form this conclusion, when obeying the NPR to plot, we discover that the left five models have a high degeneracy with each other and the base cosmology scenario, namely, these models can not be distinguished at the present stage. Note that this result is completely different form that obtained from Figure. \ref{2}.

Using the statefinder hierarchy $\{S_5^{(1)},S_5^{(2)}\}$ and obeying the NPR (see Figure. \ref{4}), we find that the evolutional trajectory of the model FMPC still deviates the base cosmology scenario and other models too much, and the models FEC and EC can be hardly distinguished from each other and the base cosmology model at the present epoch. Additionally, in Figure. \ref{4}, since the models FOC, OC and MPC have a substantially high degeneracy with the base cosmology model, one can not differentiate them from each other clearly. Furthermore, it is very constructive and interesting to use the NPR to plot about the models OC, MPC and EC (see Figure. \ref{5}), and we find that the models MPC and EC can be well distinguished form the base cosmology model and one from the other at the present epoch, however, the model OC still share a high degeneracy with the base cosmology scenario. Attractively, we can still not find out the evolutional trajectory of the model FOC. Thus, we continue to use the NPR and make a plot for the model FOC (see Figure. \ref{6}), and discover that this model exhibit a high frequency oscillating behavior in the past and approach the base cosmology scenario gradually in the future. It is worth noticing that in the plane of $\{S_5^{(1)},S_5^{(2)}\}$, we can get more detailed information by hierarchically applying the NPR three times into discriminating these models from each other.

From Figure. \ref{7}, one can easily obtain the conclusion that the models FMPC, MPC, FEC and EC can be well distinguished from the standard cosmological model and one from the other at the present stage, however, the models FOC and OC are hardly discriminated from each other and from the base cosmology scenario. In Figure. \ref{8}, one can get the same result as Figure. \ref{7}, except for the conclusion that the model MPC like the models FOC and OC can be hardly distinguished from each other and from the base cosmology scenario at the present epoch.
Note that in Figures. \ref{7} and \ref{8}, if continuing to apply the NPR for the model FOC, we will get more useful information including the evolutional trajectory of the model FOC and the fate of the universe in the planes of $\{S_3^{(2)},S_4^{(2)}\}$ and $\{S_4^{(2)},S_5^{(2)}\}$.

In Figure. \ref{5}, one may find that the evolutional trajectories of the models MPC and EC go through the $\Lambda$CDM model several times, thus, we think it is constructive to analyze the reason of the phenomenon by investigating the evolutional behaviors of the models in the plane of $\{\Omega_{m},S_3^{(1)}\}$. In Figure. \ref{9}, it is easy to be seen that all the models can be well distinguished from the $\Lambda$CDM model and one from the other at the present epoch. At the same time, one can find that the models MPC, FEC and EC cross the horizontal line more than one time, thus, the above-mentioned phenomenon can be explained well. In addition, one can discover that the model FMPC still exhibits an exotic behavior so as to be removed from the alternatives of the dark energy phenomenon without doubt.

In Figure. \ref{10}, the evolutional behaviors of the equation of state parameters for all the Cardassian models are depicted. One can find that the model EC exhibits a quintessence-like behavior, to the opposite, the model MPC exhibits a phantom-like behavior all the way, and the model FEC exhibits a quintom-like behavior. At the same time, one can discover that the models FOC and OC accord with the $\Lambda$CDM model substantially well during the evolutional processes, and the model FMPC still deviates from the standard cosmological model too much.

In Figure. \ref{11}, we have exhibited the evolutional behavior of the fractional growth parameter $\epsilon(z)$ for the aforementioned Cardassian cosmological models and the $\Lambda$CDM scenario. It is not difficult to be seen that these models, apart from the model FMPC, run closely to the standard cosmological model in the remote past. Furthermore, one can discover that there exists a high degeneracy among the models FOC, OC and FMPC at the present epoch as well as in the far future. In addition, the model FEC is well consistent with the $\Lambda$CDM scenario in a substantially long period and deviates it gradually from the present stage.

In Figure. \ref{12}, we have plotted the evolutional trajectories of all the models in the plane of $\{\epsilon(z),S_3^{(1)}\}$ by using the CND. It is worth noting that, in this figure we just use the NPR to make a plot for the models apart from the model FMPC, and adopt the distinct colorful lines for all the models which are different from those used in the previous figures. We find that the models FOC, MPC, FEC and EC can be well distinguished from the $\Lambda$CDM scenario and one from the other at the present epoch, and the model OC shares a high degeneracy with the standard cosmological model. Besides, the evolutional trajectory of the model FMPC is shown in the lower right corner of Figure. \ref{12}, and this model still has a large deviation from the $\Lambda$CDM scenario and other Cardassian models.

Subsequently, in Figure. \ref{13}, we take the $Om(z)$ diagnostic to distinguish the Cardassian cosmological models from the $\Lambda$CDM model, and one from other. It is obvious that one can find that apart from the model FMPC, the left five models share a substantially high degeneracy with each other and the $\Lambda$CDM model. Furthermore, if using the NPR to plot for these five models, we can get more detailed information. In Figure. \ref{14}, the models FOC, OC, MPC, FEC and EC can not be well distinguished from each other when $z\simeq1$, i.e., these five models share a substantially high degeneracy with each other around the present epoch. Realistically, comparing with the astronomical observations, one can not discriminate the Cardassian scenarios and the $\Lambda$CDM scenario from each other in the plane of $\{z,Om(z)\}$ at the present stage in terms of $68.3\%$ confidence level.
\section{The discussions and conclusions}
Since the elegant discovery that the universe is undergoing a phase of accelerated expansion, cosmologists have proposed a large number of models to explain the accelerated mechanism. Therefore, it is worth investigating the relationships among various kinds of cosmological models very much. In the literature, Sahni et al. \cite{37,38,39} have constructed a series of null diagnostics to distinguish different dark energy scenarios from the base cosmology scenario and one from the other, including the original statefinder, the statefinder hierarchy, the CND, the $Om(z)$ and $Om3(z)$ diagnostics, etc.

In the present paper, above all, we place constraints on the six Cardassian cosmological models which is listed in Table. \ref{tab1}, using the SNe Ia, BAO, CMB, OHD data-sets as well as the single data point from the newest event GW150914. Subsequently, we have taken the the statefinder hierarchy, the CND and the $Om(z)$ diagnostic to distinguish the Cardassian scenarios from the base cosmology scenario, and one form the other. As described in the previous section, we have plotted the evolutional trajectories of these models in the planes of $\{S_3^{(1)},S_4^{(1)}\}$, $\{S_4,S_4^{(2)}\}$, $\{S_3^{(2)},S_4^{(2)}\}$, $\{S_5^{(1)},S_5^{(2)}\}$, $\{S_3^{(2)},S_4^{(2)}\}$, $\{S_4^{(2)},S_5^{(2)}\}$, $\{\Omega_m,S_3^{(1)}\}$,$\{z,\omega\}$ $\{\epsilon(z),S_3^{(1)}\}$, $\{z,Om\}$, etc. Through deeper analysis, we have proposed the NPR for plotting in order to obtain more useful information about the relationships among different cosmological models. For a concrete instance, we find that the models FOC and OC just can be distinguished in the plane of $\{\Omega_m,S_3^{(1)}\}$ at the present epoch, however, if applying the NPR to plot for the fewer models than the six Cardassian models in the planes of $\{S_3^{(1)},S_4^{(1)}\}$, $\{S_5^{(1)},S_5^{(2)}\}$, $\{S_3^{(2)},S_4^{(2)}\}$ and $\{\epsilon(z),S_3^{(1)}\}$, we discover that the two models and the base cosmology model can be discriminated better form each other at the present stage. To be more precise, changing the plotting scale, we find that even if we can not distinguish different cosmological models through making one plot by using some null diagnostic, we can always use the NPR to make several figures hierarchically for fewer models in order to obtain the detailed information and discriminate these models better. Thus, combining several figures for the same diagnostic together, one can obtain more useful and detailed information about the relations among different models. Subsequently, to understand the phenomenon that models MPC and EC go through the fixed point $\{1,0\}$ which corresponds to the base cosmology scenario more than one time in the plane of $\{S_5^{(1)},S_5^{(2)}\}$ better, we have plotted the evolutional trajectories of
the Cardassian models and the base cosmology model in the plane of $\{\Omega_m, S_3^{(1)}\}$. It is not difficult to find that in the aforementioned figure, the evolutional trajectory of the Cardassian scenarios go through the horizontal line which corresponds to the $\Lambda$CDM scenario more than one time. In Figure. \ref{12}, by adopting the CND $\{\epsilon(z),S_3^{(1)}\}$, one can easily find that the models FOC, MPC, FEC and EC can be well distinguished from the base cosmology scenario and one from the other at the present epoch, and the model OC shares a high degeneracy with the standard cosmological model. In addition, as a supplement, using the $Om(z)$ diagnostic, we find that apart from the model FMPC, the left five Cardassian models share a substantially high degeneracy with each other and the base cosmology model. However, if taking the NPR to make a plot for the models FOC, OC, MPC, FEC and EC, we discover that the five models can not be well discriminated from each other when $z\simeq1$, i.e., these five models share a substantially high degeneracy with each other around the present epoch. Actually, comparing with the astronomical observations, one can not distinguish the Cardassian scenarios and the base cosmology scenario from each other in the plane of $\{z,Om(z)\}$ at the present epoch in terms of $68.3\%$ confidence range. Interestingly and importantly, it is noteworthy that, from the planes of $\{S_4,S_4^{(2)}\}$, $\{S_5^{(1)},S_5^{(2)}\}$, $\{S_3^{(2)},S_4^{(2)}\}$, $\{\Omega_m,S_3^{(1)}\}$,$\{z,\omega\}$ $\{\epsilon(z),S_3^{(1)}\}$ and $\{z,Om\}$,
we find that the model FMPC can be directly removed from the possible candidates of dark energy phenomenon, since its evolutional behavior deviates badly from the standard cosmological scenario.

The most interesting point in this situation is that, specifically, we introduce the NPR to plot hierarchically for the Cardassian cosmological models, so we can obtain more detailed information in order to distinguish these models from each other and the base cosmology model better. Generally speaking, the larger the number of the models are, the more effective and powerful plotting method based on the NPR becomes. Moreover, if adopting the NPR to plot hierarchically for several cosmological models by using only one null diagnostic, one can usually discriminate these models well from each other since the detailed information from different hierarchical figures has been obtained. Therefore, there is no need to use many geometrical null diagnostics to discriminate different cosmological models as before, namely, we reduce the necessity to adopt different diagnostics and the number of diagnostics to distinguish different models from each other by using the plotting method based on the NPR.
It is worth noticing that, utilizing the NPR to plot hierarchically for some models, if one still can not distinguish at least two models from each other through taking one null diagnostic, then follow a rational line, it is necessary to adopt other geometrical diagnostic to expect for discriminating the corresponding models better. Hence, obviously, we avoid the blindness to use different diagnostics when discovering that some models can be hardly differentiated from each other and develop a logical line to adopt different diagnostics which is based on the NPR.

\acknowledgments
During the preparation process of the present work, we thank Professors Bharat Ratra and Saibal Ray for helpful correspondences and interesting discussions on gravitational wave physics in cosmology and wormhole astrophysics. The author Deng Wang thank Prof. Jing-Ling Chen for constructive communications about quantum information in general relativity. This work is supported in part by the National Science Foundation of China.

\end{document}